\documentclass[prd,showpacs,preprint,eqsecnum,amsmath,amssymb,nofootinbib]{revtex4}
\usepackage{graphicx, tikz}
\usepackage{subcaption}
\usepackage{amssymb,amsmath, mathrsfs}
\usepackage{amssymb, graphics, setspace}
\usepackage{hyperref}
\usepackage[all]{xy}
\newcommand{\be}{\begin{equation}}
\newcommand{\ee}{\end{equation}}
\newcommand{\bea}{\begin{eqnarray}}
\newcommand{\eea}{\end{eqnarray}}
\newcommand{\bes}{\begin{subequations}}
\newcommand{\ees}{\end{subequations}}

\begin{document}

\title{Quantum correlations across the horizon\\ in acoustic and gravitational Black Holes}
\author{Roberto~Balbinot}
\email{balbinot@bo.infn.it}
\affiliation{Dipartimento di Fisica dell'Universit\`a di Bologna and INFN sezione di Bologna, Via Irnerio 46, 40126 Bologna, Italy
%
}
\author{Alessandro~Fabbri}
\email{afabbri@ific.uv.es}
\affiliation{Departamento de F\'isica Te\'orica and IFIC, Universidad de Valencia-CSIC, C. Dr. Moliner 50, 46100 Burjassot, Spain
}

\bigskip\bigskip

\begin{abstract}
We investigate, within the framework of Quantum Field Theory in curved space, the correlations across the horizon of a black hole in order to highlight the particle-partner pair creation mechanism at the origin of Hawking radiation. The analysis concerns both acoustic black holes, formed by Bose-Einstein condensates, and gravitational black holes. More precisely, we have considered a typical acoustic black hole metric with two asymptotic homogeneous regions and the Schwarzschild metric as describing a gravitational black hole. By considering equal time correlation functions, we find a striking disagreement between the two cases: the expected characteristic peak centered along the trajectories of the Hawking particles and their partners seems to appear only for the acoustic black hole and not for the gravitational Schwarzschild one.
The reason for that is the existence of a “quantum atmosphere" displaced from the horizon as the locus 
of origin of Hawking radiation together, and this is the crucial aspect, with the presence of a central singularity in the gravitational case swallowing everything is trapped inside the horizon.
Correlations however are not absent in the gravitational case: to see them one has simply to consider correlation functions at unequal times which indeed display the expected peak.

\end{abstract}
\maketitle
\section{Introduction}

One of the most amazing properties of black holes (BHs) is that, despite their name, they are not really black as they emit thermal radiation. This is the spectacular result reached by Hawking in 1974 \cite{hawking}, a milestone of modern theoretical physics. Unfortunately this radiation is so weak that it is almost impossible (at least so far) to reveal it. More precisely, the emission temperature for a solar mass BH is of order $10^{-7} K$ (which should be compared to the $2.7\ K$ of the cosmic microwave background) and even lower for more massive BHs since this temperature scales inversely proportional to the mass. The only hope could come from primordial BHs \cite{carr} whose mass can be significantly lower than the solar mass., but so far no evidence of such kind of signals has been revealed (see for example \cite{macgibbon}).

The mechanism responsible for the Hawking radiation is intrinsically quantum mechanical, namely the conversion of quantum vacuum fluctuations in on shell real particles. More precisely, pairs of correlated particles are created, one member outside the BH horizon escaping to infinity and constituting the thermal radiation, the other, called partner \cite{mapa}, inside the horizon getting swallowed by the interior singularity. Note that particle and partner have opposite Killing energy and this allows the mechanism of pair creation to work even in static or stationary space-times.

In 1981 Unruh \cite{Unruh} showed that the same quantum mechanism could work in a fluid crossing the speed of sound, opening the possibility of observing the analog of Hawking radiation in systems less exotic than BHs and manageable in a laboratory. This started the interest in the so called analog models, condensed matter systems which mimic characteristic features of BHs and the early Universe and which can be used as a “laboratory'' to test the predictions of Quantum Field Theory (QFT) in curved space-times  \cite{Barcelo:2005fc}. 
Most of the research has been concentrated to sonic analog of BHs constructed from Bose-Einstein condensates (BECs) \cite{gacz} where the expected Hawking temperature  $\sim 10\ nK$ is only one order of magnitude less than the background ($\sim 100\ nK$). Nevertheless competing effects like thermal emission can still cover the Hawking radiation. To overcome this, it was realized that Hawking radiation has a characteristic imprint on the equal time density-density correlation function when one point is taken outside the sonic horizon and the other inside \cite{paper1, cfrbf}. Once the horizon is formed and a stationary regime is reached, a well defined peak appears reflecting the correlation between the Hawking particles and their partners. This is the smoking gun of the presence of Hawking radiation. Using this input J. Steinhauer and co-workers by a series of experiments were finally able to reveal it \cite{jeff2016, jeff2019}.

In recent time the attention of the experimentalists has been moved from the stationary regime, which is theoretically almost understood, to the dynamic formation of time dependent analog BHs horizons, like the experiment reported by Steinhauer's group \cite{jeff-nuovo}. They were able to follow the ramp-up of Hawking radiation following the formation of a sonic horizon in a BEC looking at the time evolution of  the in-out density correlation function.
In this paper we will show how, using the powerful methods of QFT in curved space-times, one can theoretically describe this ramp-up. This is done in a simple toy model of the sonic BH formation which allows an analytical treatment able to reproduce the qualitative features of the experiment \cite{nostro-nuovo}. The temporal  evolution of the in-out correlation function towards its stationary configuration we obtain will shade light to when and where Hawking radiation emerges out of the vacuum fluctuations.

Locating the origin of Hawking radiation in a BH is a long standing issue. It was first discussed by Unruh \cite{unruh1977} analyzing the propagation of the modes associated to a quantum field in a BH spacetime. Later Giddings \cite{qatm1}  introduced the idea of a “quantum atmosphere” significantly displaced from the horizon as the locus where Hawking radiation emerges out of vacuum fluctuations. This work was followed by other authors refining the analysis \cite{qatm2, qatm3}. The results of these studies locate in a Schwarzschild black hole the region where the Hawking particles emerge at a distance $O(1/\kappa)$ from the horizon, where $\kappa$ is the horizon's surface gravity. 
We will see that our analysis based on correlation across the horizon will give strong support on the existence of a quantum atmosphere also for acoustic BHs.

Usually the studies on Hawking radiation for gravitational BHs focus on the region exterior to the horizon. One looks for example to the particle spectrum revealed far away from the BH analyzing the modes of the quantum field associated to the emitted particles and the related Bogoliubov transformations \cite{hawking}. At a deeper level, one tries to get information on Hawking radiation by studying the renormalized stress energy tensor of the quantum field \cite{dfu}.

Here, inspired by the work on acoustic BHs, we will also study the particle-partner correlation across the horizon in Hawking radiation for a BH formed by gravitational collapse.
Of course, one has no access to the region inside the horizon since this surface is now a causal boundary preventing information from inside to leak outside. So this study has just a theoretical interest. Nevertheless the complete different pattern of the equal time in-out correlations which will emerge compared to the ones in acoustic BHs because of the presence of an interior singularity will help to understand better the role of the quantum atmosphere in the Hawking effect. 

The paper is organized as follow. In Section II we discuss the correlations across the horizon of the density fluctuations in an acoustic BH formed by a BEC. The toy model used to mimic the horizon formation and the technical details are left in Appendix A.
In Section III the same kind of analysis is performed for a BH formed by gravitational collapse of a null shell. Some mathematical details are given in Appendix B.
Section IV is devoted to the conclusions.

\section{In-out correlations in BEC sonic BHs}
\label{s2}

In Ref. \cite{nostro-nuovo}  we have shown, using a simple model, how, with the methods of QFT in curved space,  
one can describe quite well the time evolution of the density-density correlation function in a BEC once a sonic horizon is formed (see also \cite{renaud2010, mcp}). The goal was to locate where and when the signal of the presence of Hawking radiation appears.

In this paragraph we review the model and discuss the results. The technical details are given in Appendix \ref{appendixA}. 

The model we use to mimic the BH formation is quite simple so that a complete analytical treatment is possible. However we are confident that its outcomes describe, at least qualitatively, correctly the physics of the ramp-up of Hawking radiation towards the stationary regime characterized by the peak in the correlation function mentioned in the Introduction. 

We assume a one-dimensional BEC flow, directed from right to left along the $x$ axis at a constant velocity $V\ (<0)$; the density $n$ of the condensate is also constant. By properly modulating the speed of sound $c$ we can generate the formation of an acoustic BH. To this end the profile of $c$ is assumed to vary in time as following
\bea \label{dueuno}
\begin{cases}
c &= c_{in} , \ \ \   t<0   \\              
c &= |V|\left( 1+\gamma \tanh \frac{\kappa x}{\gamma|V|}\right)            ,\ \ \    t>0\\
\end{cases}
\eea
where $c_{in}\ (>|V|)$ is constant like $\kappa$ and $\gamma=\frac{2}{3}$. The chosen profile describes an uniform subsonic flow for $t<0$. At $t=0$ a sonic BH forms: the region $x>0$ remains subsonic, while for $x<0$ the flow is supersonic. The sonic horizon is located at $x=0$ and $\kappa$ is its surface gravity. We remark that modulations of the speed of sound in a BEC, even suddenly, are now obtained with routine procedures in laboratories: for example using Feshbach resonances to vary the atom-atom interaction coupling \cite{pi-st}. 

The equal time correlator $G_2^{(1)}(t;x,x')$ of the 1D density fluctuation operator can be approximated within the gravitational analogy using the methods of QFT in curved space as following \cite{paper1}
\be G_2^{(1)}(t;x,x') =\sqrt{\frac{n(x)n(x')}{m^2c^3(x)c^3(x')}}  \lim_{t\to t'} D  \langle  \delta \hat \theta^{(2)}  \delta \hat \theta^{(2)} \rangle 
  \label{duedue}   \ee
  where $n(x)$ is the 1D density of the condensate (constant in our model), $m$ the mass of the single atom of the BEC, $D$ the differential operator 
 \be D=(\partial_t+V\partial_x)(\partial_{t'}+V\partial_{x'}) \label{duetre} \ee  and $\langle \delta \hat \theta^{(2)}(t,x) \delta \hat \theta^{(2)}(t',x')  \rangle$ is the two-point function 
 of a $1+1$ dimensional massless scalar field  $\delta \hat \theta^{(2)}$, whose field equation is  
 \be \hat \Box  \delta \hat \theta^{(2)}  =0 \  , \label{duequattro}\ee
 propagating in the $1+1D$ acoustic metric
 \be ds^2 =  -( c^2 - V^2 ) dt^2  - 2Vdtdx +dx^2      \label{duecinque}\ee
and $\hat \Box$ is the covariant D'Alembert operator computed from the above metric.  

\begin{figure}[h]
\centering
\begin{subfigure}[h]{0.45\textwidth}
{\includegraphics[width=2in]{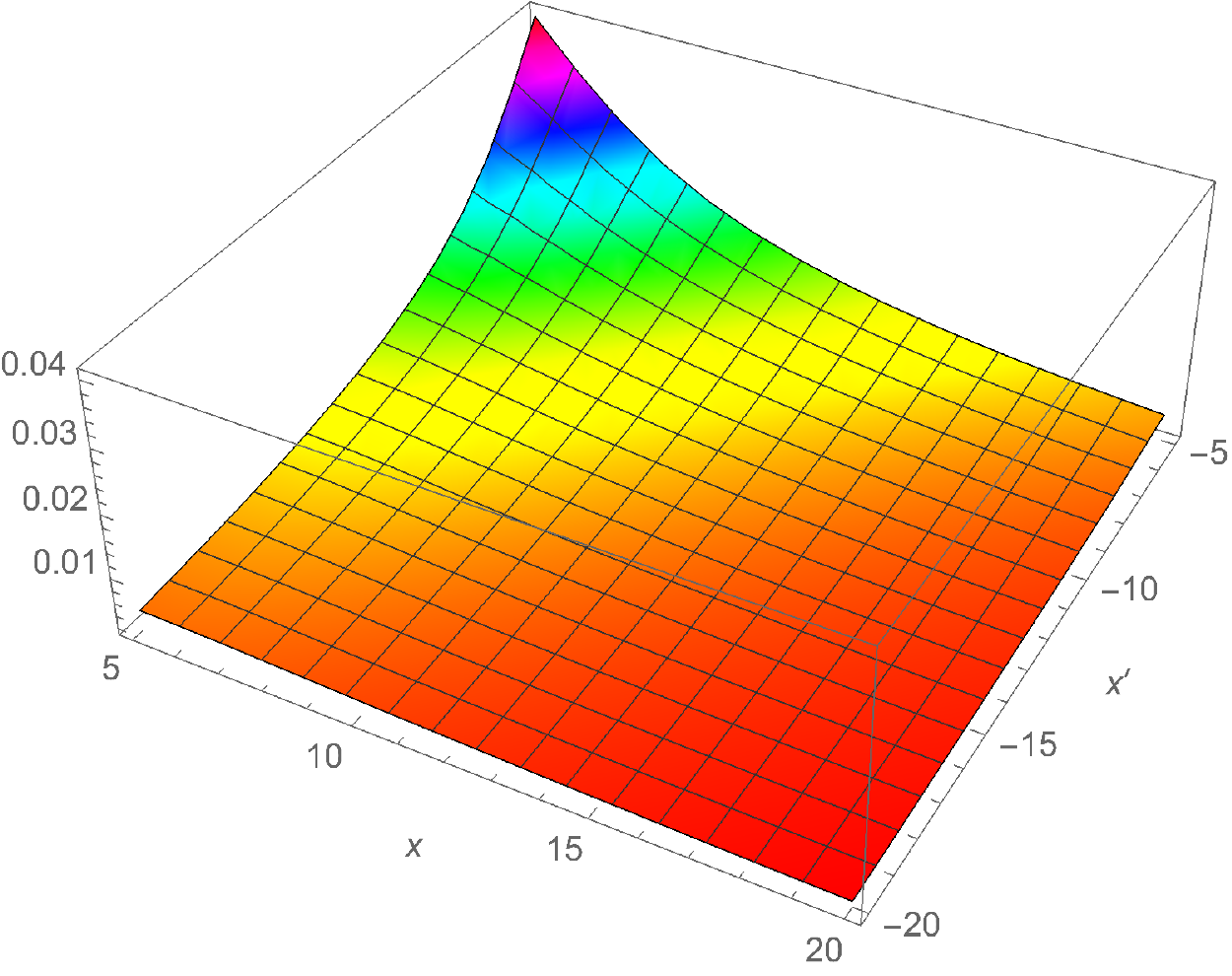}}
\caption{}
\label{1a}
\end{subfigure}
\begin{subfigure}[h]{0.45\textwidth}
{ \includegraphics[width=2in]{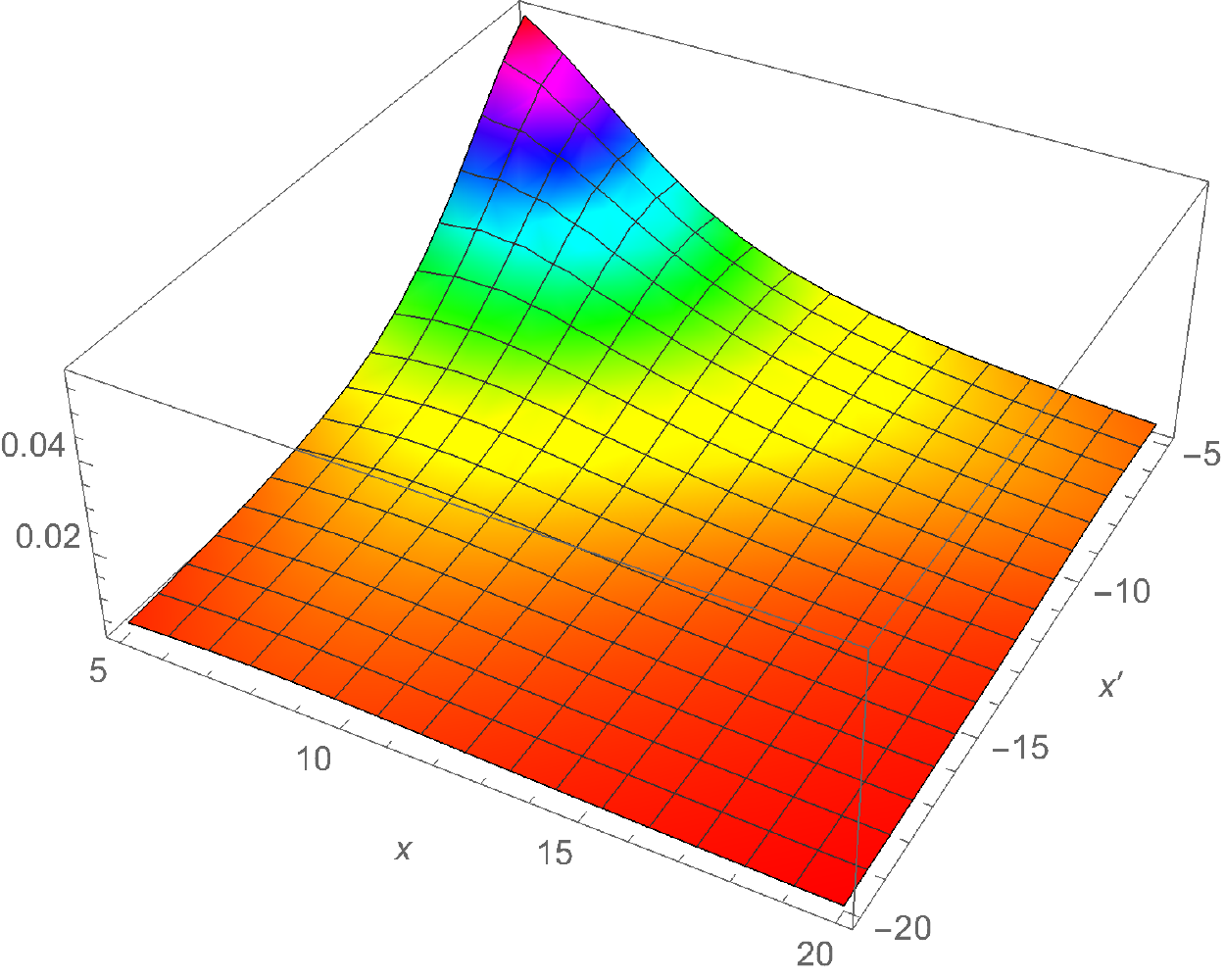}}
\caption{}
\label{1b}
\end{subfigure}
\begin{subfigure}[h]{0.45\textwidth}
{\includegraphics[width=2in]{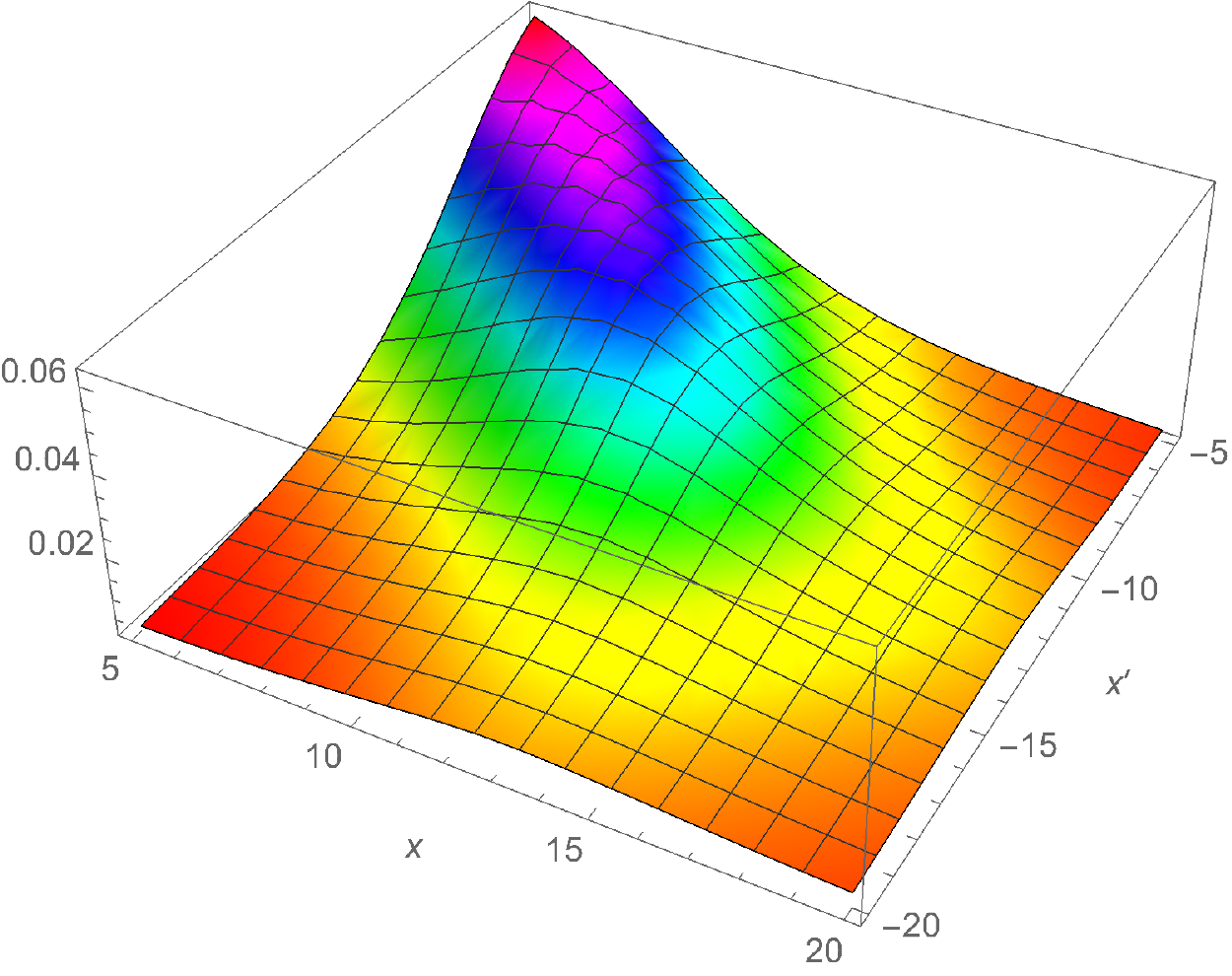}}
\caption{}
\label{1c}
\end{subfigure}
\begin{subfigure}[h]{0.45\textwidth}
{\includegraphics[width=2in]{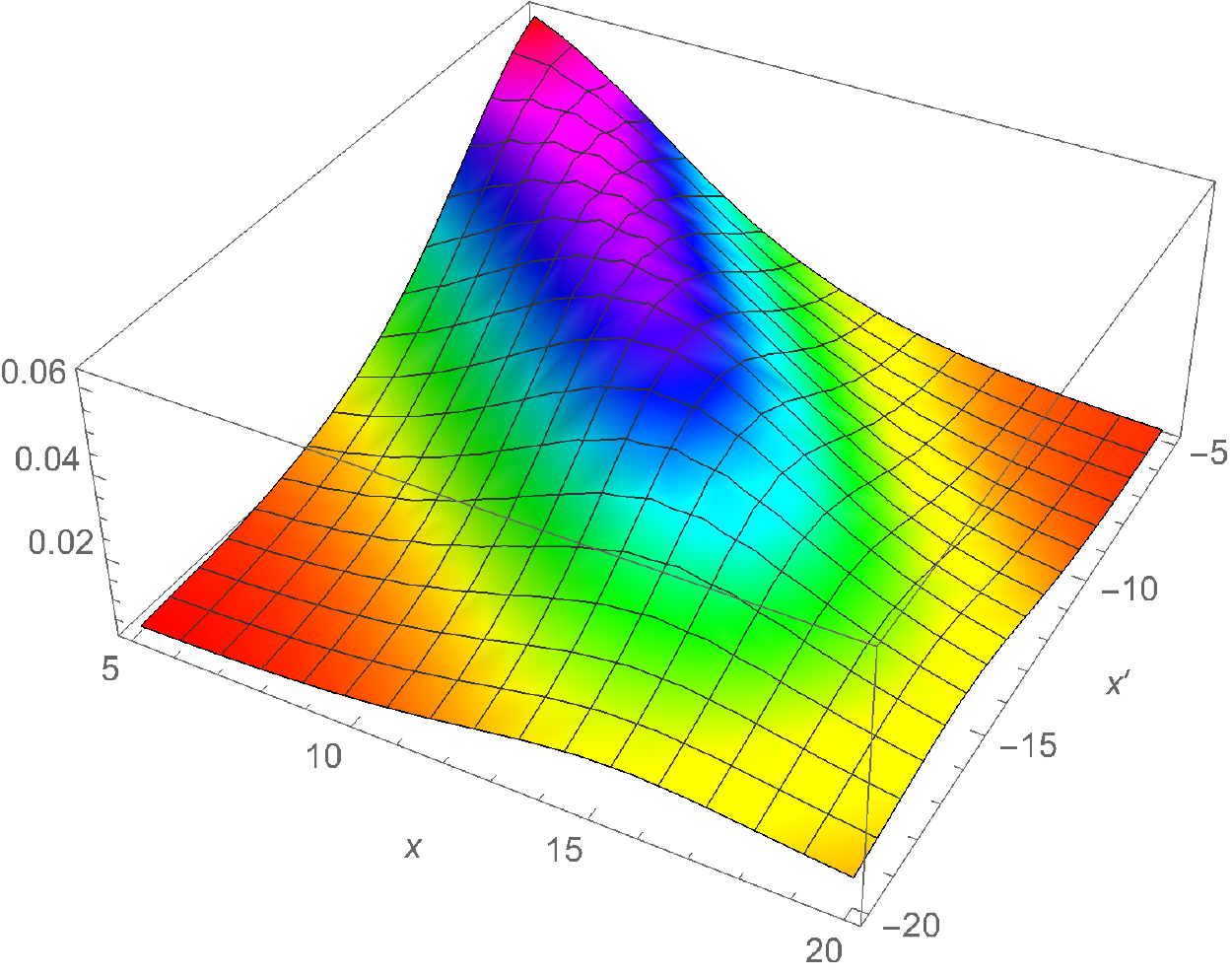}}\quad\quad
\caption{}
\label{1d}
\end{subfigure}
\caption{\label{fig1}  Plots of the absolute value of  $G_2^{(1)}(t;x,x')$, for $x>0$ (outside the horizon) and $x'<0$ (inside the horizon), at four increasing times $t_1=1/\kappa$ (a), $t_2=2/\kappa$ (b), $t_3=4/\kappa$ (c), $t_4=5/\kappa$ (d). Here and in the figures that follow of this section we have plotted the correlator up to the overall factor $\frac{\hbar n}{4\pi m}$ and chosen the values $\kappa=\frac{1}{4}, |V|=1, c_{in}=\frac{3}{2}$. }
\end{figure}

This approximation neglects backscattering of the modes caused by the inhomogeneity of the BEC (i.e. curvature of the acoustic metric) \cite{paper2013}. This backscattering produces two secondary peaks in the correlation pattern beside the main one referred to in the Introduction \cite{Macher:2009nz, rpc, lrcp}, but so far no experiment has been able to see them given the much weaker signature of these ones compared to the former. The mathematical  details of our construction are given in Appendix \ref{appendixA}. Here we discuss the results we have obtained. In Fig. \ref{fig1} we have plotted  $G_2^{(1)}(t;x,x')$ for points $x>0$ (outside the horizon) and $x'<0$ (inside the horizon) at four increasing times $t_1=1/\kappa,\ t_2=2/\kappa,\ t_3=4/\kappa,\ t_4=5/\kappa$. 
      
\begin{figure}[h]
\includegraphics[width=3in]{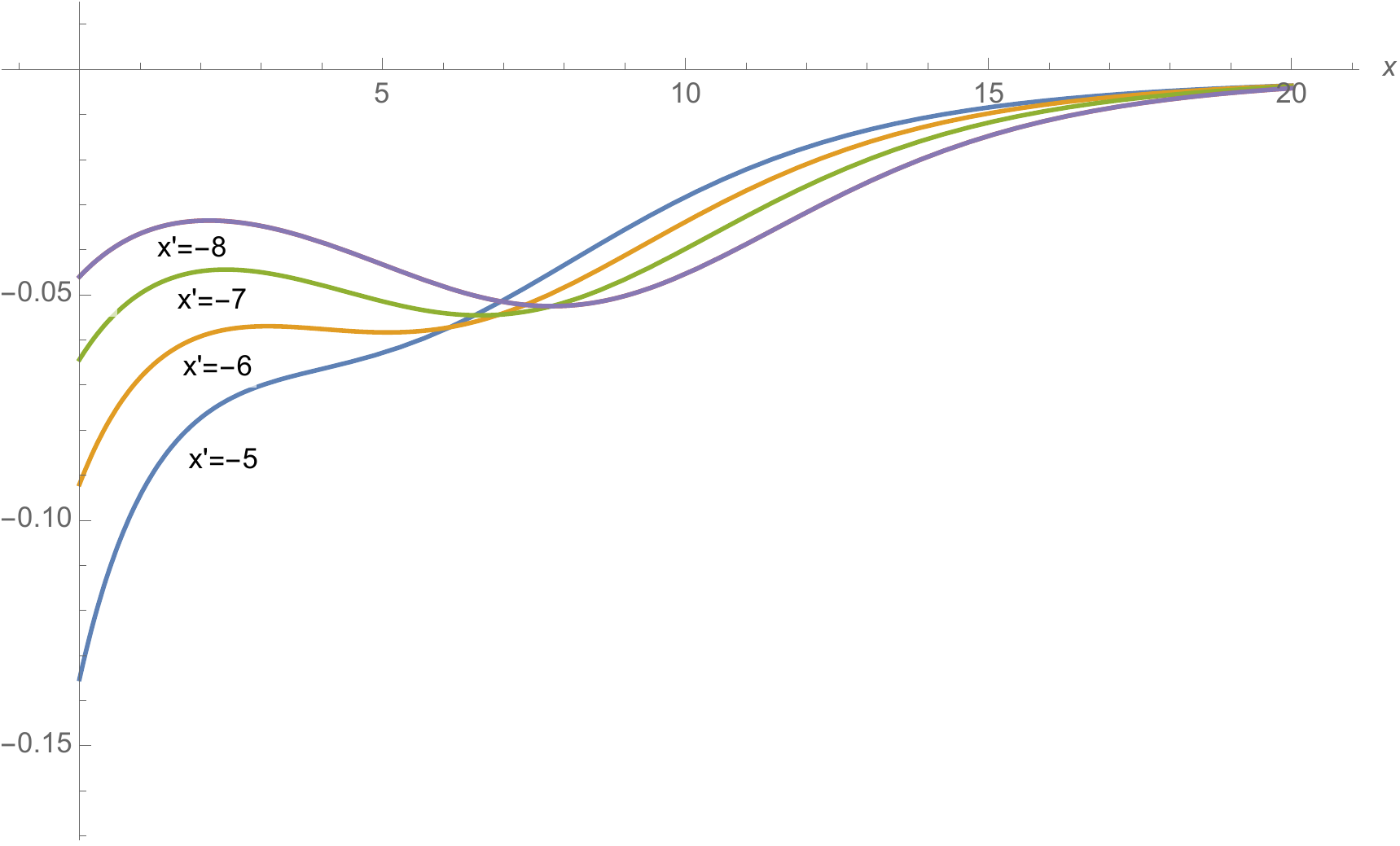} 
\caption{\label{fig2} Plot of $G_2^{(1)}(t;x,x')$ at $t=\frac{10}{\kappa}$, as a function of $x$ for various (indicated) fixed $x'$.}
\end{figure}

At late time, once the stationary regime is achieved, a peak at $x'=-x$ is expected, signaling the correlation of the Hawking particles and their partners. The signal is the one that has been experimentally observed \cite{jeff2016, jeff2019, jeff-nuovo}. In our model, the late time limit is governed by the condition $e^{-\kappa t}\sinh \frac{3\kappa |x|}{2|V|}=cst\ll1$ (see eq. (\ref{aventuno})). Note that this condition is reached earlier in the near-horizon region and then it spreads out. From Fig. \ref{fig1}, where we have plotted the absolute value of
$G_2^{(1)}(t;x,x')$ (the correlator is negative), one sees indeed the formation of a peak located at $x'=-x$.
Note however that  this holds for $x=-x'$ sufficiently far away from the horizon. To see this more clearly,
in Fig. (\ref{fig2}) we plot the correlator evaluated at  $t=\frac{10}{\kappa}$ as a function of $x$ for various fixed points $x'$: $x'=-5,-6,-7,-8$. The peak is located at $x=-x'$ only for $|x'|\gtrsim 7|V|/4\kappa$. For values of $x'$ closer to the horizon the peak does not appear. So there is no sign of correlations in the near horizon region. 

Let us understand the reason. The correlator is given (see eq. (\ref{adiciotto})) by

\bea
 && G_2^{(1)}(t;x, x')
 = -\frac{\hbar n}{4\pi  mc(x)^{1/2}c(x')^{1/2}}
  \Big[ \frac{1}{(c(x)-|V|)(c(x')-|V|)} \frac{du_{in}}{du}\frac{du'_{in}}{du'}  \frac{1}{(u_{in}-u'_{in})^2} +\nonumber \\ && \frac{1}{(c(x)+|V|)(c(x')+|V|)}  \frac{dv_{in}}{dv} \frac{dv'_{in}}{dv'} \frac{1}{(v_{in}-v'_{in})^2}  \Big] |_{t=t'}\equiv  G_{2,u}^{(1)}(t;x, x') +G_{2,v}^{(1)}(t;x, x') \ . \label{duesei} \eea

The relevant term is the first one in eq. (\ref{duesei}) describing the correlations of the positive (Killing) energy Hawking particles and their negative (Killing) energy partners. Let us call this term 
$G_{2,u}^{(1)}(t;x, x')$. The other one $G_{2,v}^{(1)}(t;x, x')$ is related to the $v$ sector. 

\begin{figure}[h]
\includegraphics[width=2in]{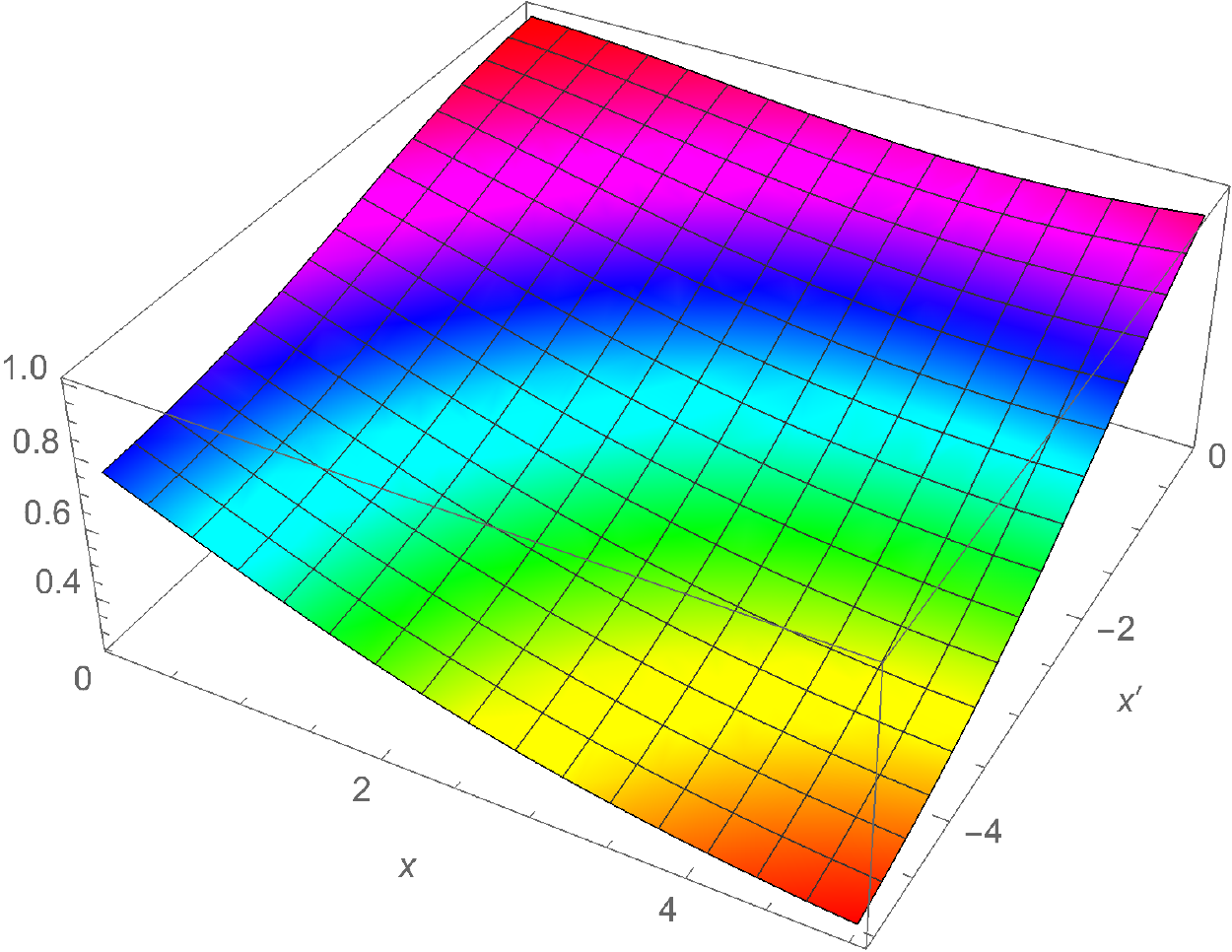} 
\caption{\label{fig3} $\frac{G_{2,v}^{(1)}(t;x,x')}{G_{2,u}^{(1)}(t;x,x')}$, defined by (\ref{duesei}), at $t=\frac{10}{\kappa}$ in the near-horizon region.}
\end{figure}

At late advanced time we have $u_{in}=\pm \frac{e^{-\kappa u}}{|A|}$ (see eq. (\ref{atredici})) where the minus sign refers to points outside the horizon ($x>0$) and the plus sign to points inside the horizon ($x<0$). In this limit we have
\be 
G_{2,u}^{(1)}(t;x, x') =-\frac{\hbar n\kappa^2}{4\pi mV^2\gamma^2\sqrt{c(x)c(x')}}\frac{\cosh\beta x\cosh\beta x'}{(\sinh\beta x-\sinh\beta x')^2}\ , \label{duesette}
\ee
where $\beta\equiv \frac{\kappa}{\gamma|V|}$. For $x$ and $x'$ far away from the horizon the absolute value of $G_{2,u}^{(1)}(t;x,x')$ has a local maximum  at $x=-x'$ while for $x,x'\to 0$ we have 
\be \label{nuova}
\lim_{x,x'\to 0} G_{2,u}^{(1)}(t;x, x') =-\frac{\hbar n}{4\pi m|V|}\frac{1}{(x-x')^2}= \lim_{x,x'\to 0} G_{2,v}^{(1)}(t;x, x')\ . \ee
Away from the horizon $G_{2,v}^{(1)}(t;x, x')$ rapidly decreases, see Fig. (\ref{fig3}), and becomes one order of magnitude smaller than $G_{2,u}^{(1)}(t;x, x')$.

From these considerations we can deduce that for $x,x'$ close to the horizon the light-cone singularity of the two-point function starts dominating and the local minimum disappears (see Fig. (\ref{fig2})) \cite{schutzholdunruh}. The correlator describes just  diverging vacuum fluctuations as $x=-x'\to 0$. So we see that the emergence of the Hawking's particle-partner pair creation out of vacuum fluctuations does not occur close to the horizon but in a region, named ``quantum atmosphere'' by Giddings \cite{qatm1}, outside the horizon which within our model can be located at $x=-x'\sim 7$ and which corresponds to a distance of  $\frac{7|V|}{4\kappa}$. One important aspect that also emerges is that the peak signal does not appear immediately after the formation of the horizon ($t=0$), but after a characteristic time of the order $4/\kappa$.

To reveal the correlations close to the horizon one needs to consider the correlator no longer at equal times. At late time we have  
\be 
G_{2,u}^{(1)}(t,t';x, x') =-\frac{\hbar n\kappa^2}{4\pi mV^2\gamma^2\sqrt{c(x)c(x')}}\frac{\cosh\beta x\cosh\beta x'}{(e^{-d}\sinh\beta x-e^{d}\sinh\beta x')^2}\ , \label{dueotto}
\ee
where $d=\frac{\kappa}{2}(t-t')\equiv  \frac{\kappa}{2}\Delta t$. 

\begin{figure}[h]
\centering
\begin{subfigure}[h]{0.45\textwidth}
{\includegraphics[width=2in]{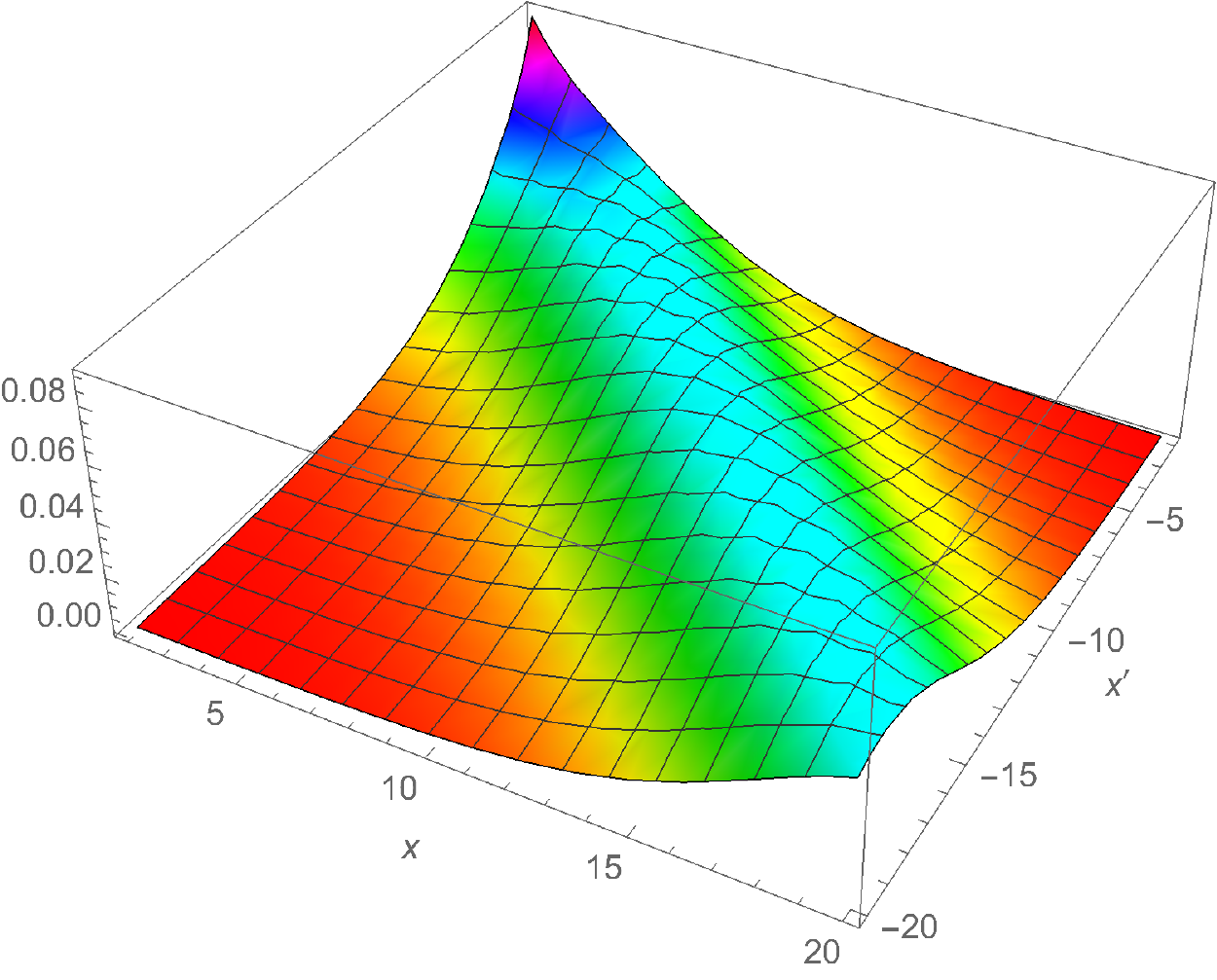}}
\caption{}
\label{4a}
\end{subfigure}\\ 
\begin{subfigure}[h]{0.45\textwidth}
{ \includegraphics[width=2in]{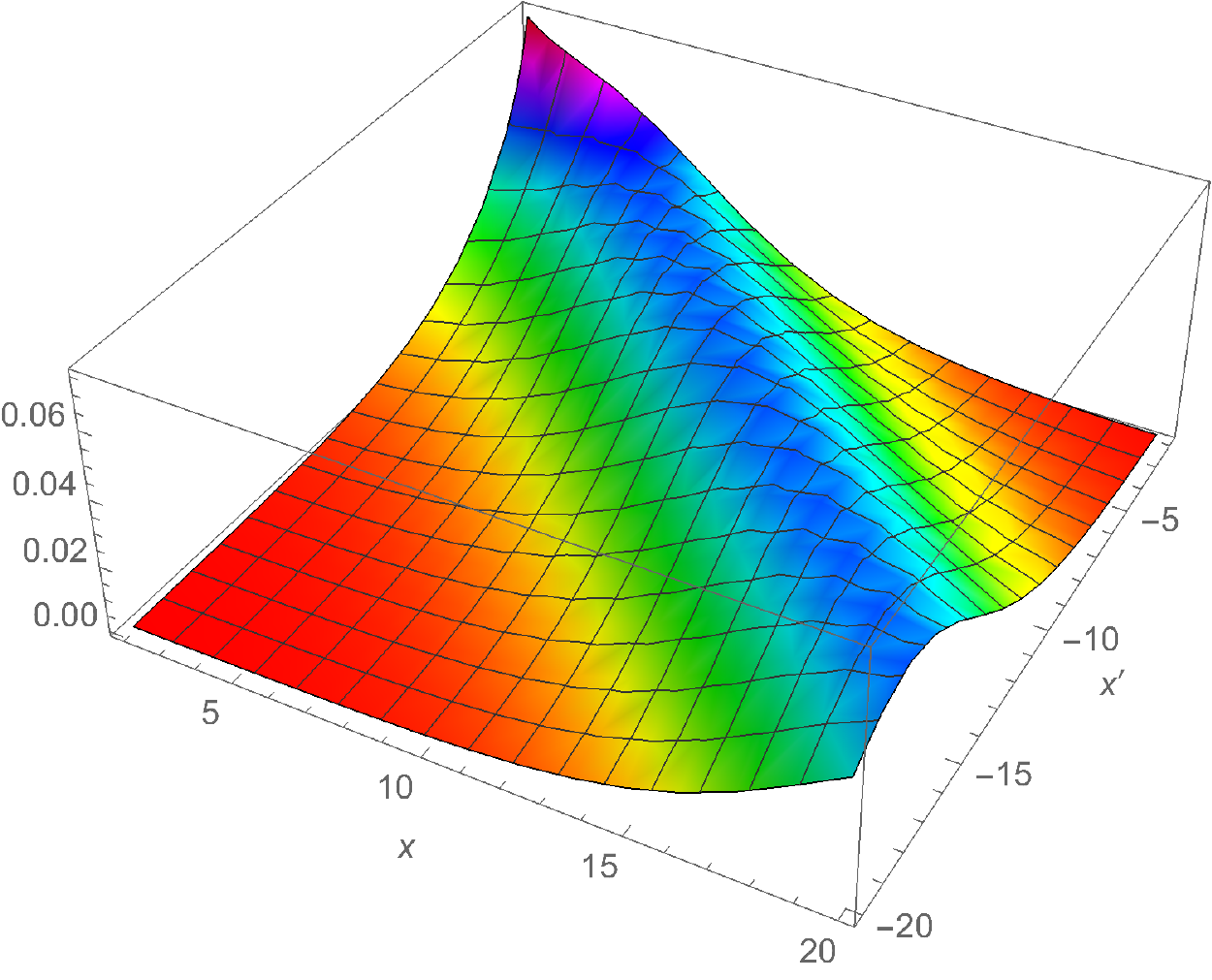}}
\caption{}
\label{4b}
\end{subfigure}\\
\begin{subfigure}[h]{0.45\textwidth}
{\includegraphics[width=2in]{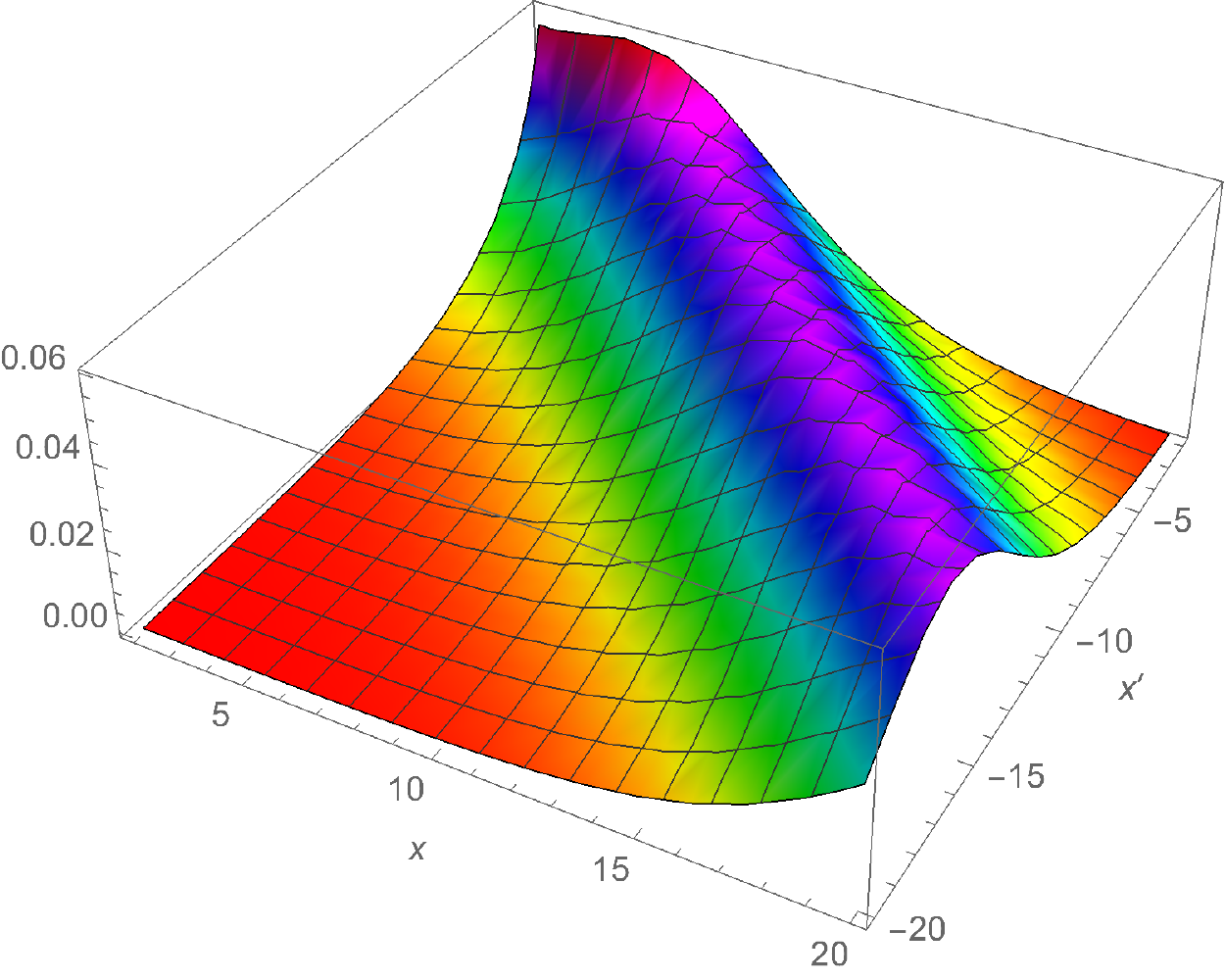}}
\caption{}
\label{4c}
\end{subfigure}
\caption{\label{fig4} Absolute value of the unequal time correlator (\ref{dueotto})  \\ at three increasing $\Delta t$ ($=1/2\kappa$ (a), $1/\kappa$ (b), $3/2\kappa$(c)) intervals.   }
\end{figure}

\begin{figure}[h]
\centering
\begin{subfigure}[h]{0.45\textwidth}
{\includegraphics[width=3in]{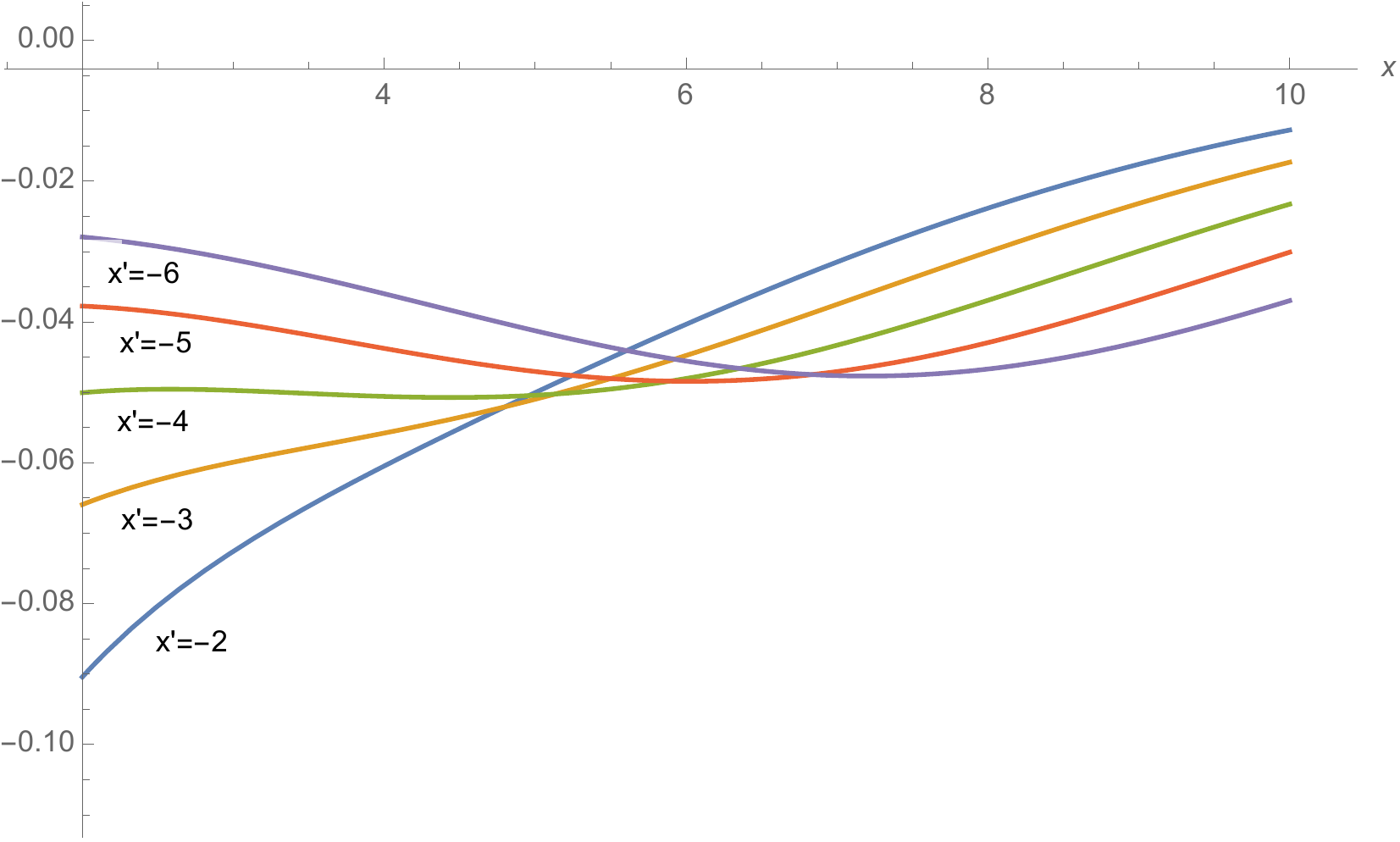}}
\caption{}
\label{5a}
\end{subfigure}\\ 
\begin{subfigure}[h]{0.45\textwidth}
{ \includegraphics[width=3in]{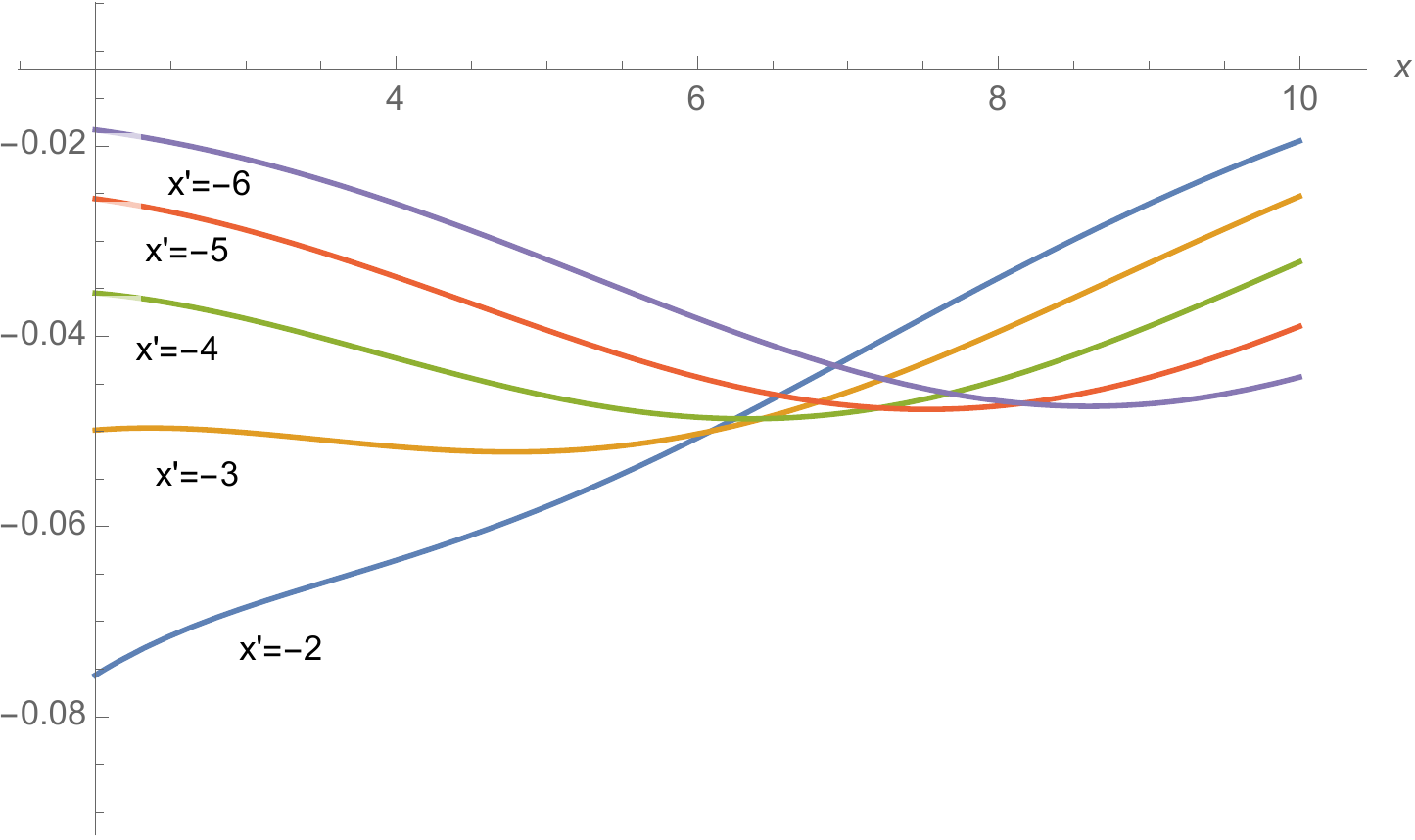}}
\caption{}
\label{5b}
\end{subfigure}\\
\begin{subfigure}[h]{0.45\textwidth}
{\includegraphics[width=3in]{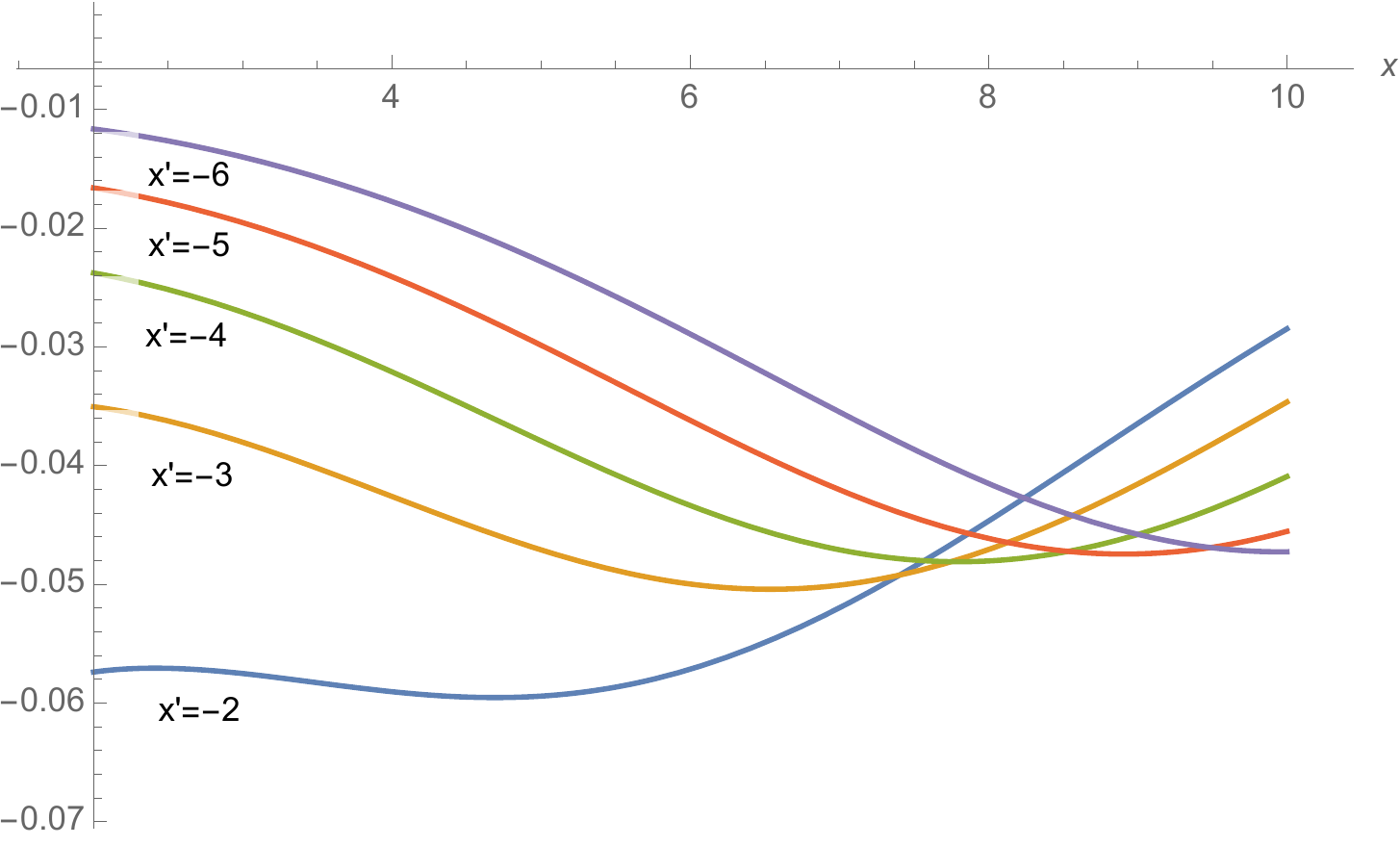}}
\caption{}
\label{5c}
\end{subfigure}
\caption{\label{fig5} Unequal time correlator (\ref{dueotto}) as a function of $x$ for five different fixed values of $x'$ at three increasing $\Delta t$ ($=1/2\kappa$ (a), $1/\kappa$ (b), $3/2\kappa$(c)) intervals.   }
\end{figure}

In Fig. (\ref{fig4}) we plot the absolute value of the unequal time correlator (\ref{dueotto}) at three increasing $\Delta t$ ($=1/2\kappa, 1/\kappa, 3/2\kappa$) intervals and, in Fig. (\ref{fig5}),  as a function $x$ for five different values of $x'$.

One sees that the values of $x'$ at which the peak appears decreases towards the horizon as $\Delta t$ increases, while for large enough $x,x'$ the peak is located at $u=u'$, i.e.
\be \label{duenove} u=t-\frac{1}{\kappa}\ln\sinh \beta x=t'-\frac{1}{\kappa}\ln\sinh|\beta x'|\ , \ee
along the trajectories of the Hawking particle and its partner as expected. 

The analysis we have performed so far considered a BEC whose temperature is zero. Experimentally this is difficult to achieve, so one should consider the case in which the condensate has an initial temperature $T\neq 0$. So for $t<0$ we have a thermal distribution of phonons characterized by an occupation number 
\be    N_{\omega_{u(v)}} =   \frac{1}{e^{\frac{\hbar\omega_{u(v)}}{k_BT}}-1}\  , \label{duedieci}    \ee
where 
\be    \omega_u = \frac{\omega c_{in}}{c_{in}-|V|}\ , \  \ \omega_v=\frac{\omega c_{in}}{c_{in}+|V|}\ , \label{dueundici} \ee    
are the Doppler rescaled frequencies. 
The corresponding equal time correlator is given now (see Appendix A) as
\bea
&& G_{2\ T}^{(1)}(t;x,x')
 = -\frac{\hbar n}{4\pi  mc(x)^{1/2}c(x')^{1/2}}
 \Big[ \frac{1}{(c(x)-|V|)(c(x')-|V|)} \frac{du_{in}}{du}\frac{du'_{in}}{du'}  \frac{A_u^2}{\sinh^2 A_u (u_{in}-u'_{in})}     \nonumber \\ &&+ \frac{1}{(c(x)+|V|)(c(x')+|V|)}  \frac{dv_{in}}{dv} \frac{dv'_{in}}{dv'}  \frac{A_v^2}{\sinh^2A_v (v_{in}-v'_{in})}   \Big] |_{t=t'}\ ,  \label{duedodici}   \eea
 where 
 \be  A_{u(v)}=\frac{\pi k_B T(c_{in}\mp |V|)}{\hbar c_{in}}\ .\ee 

\begin{figure}[h]
\centering
\begin{subfigure}[h]{0.45\textwidth}
{\includegraphics[width=2in]{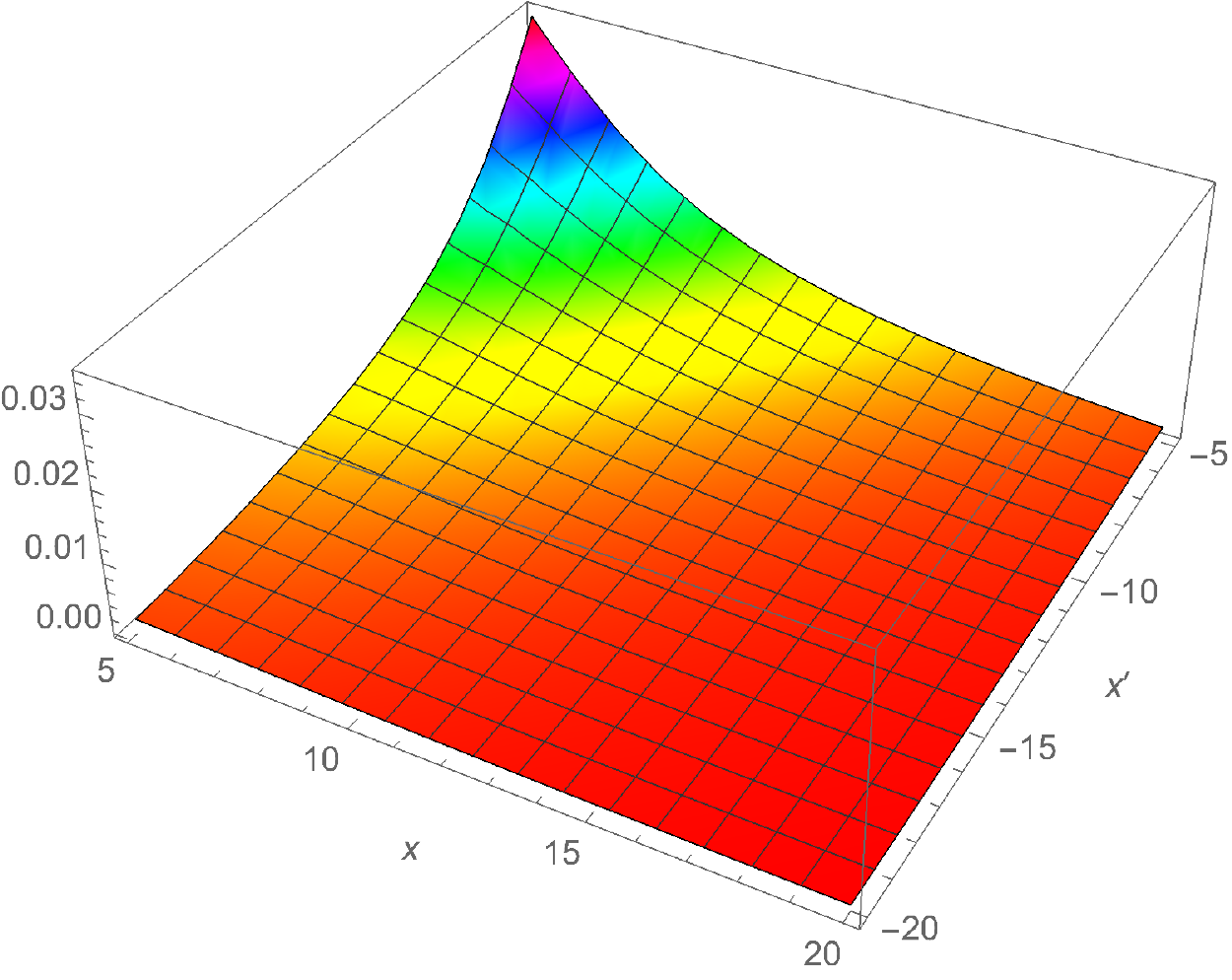}}
\caption{}
\label{6a}
\end{subfigure}
\begin{subfigure}[h]{0.45\textwidth}
{ \includegraphics[width=2in]{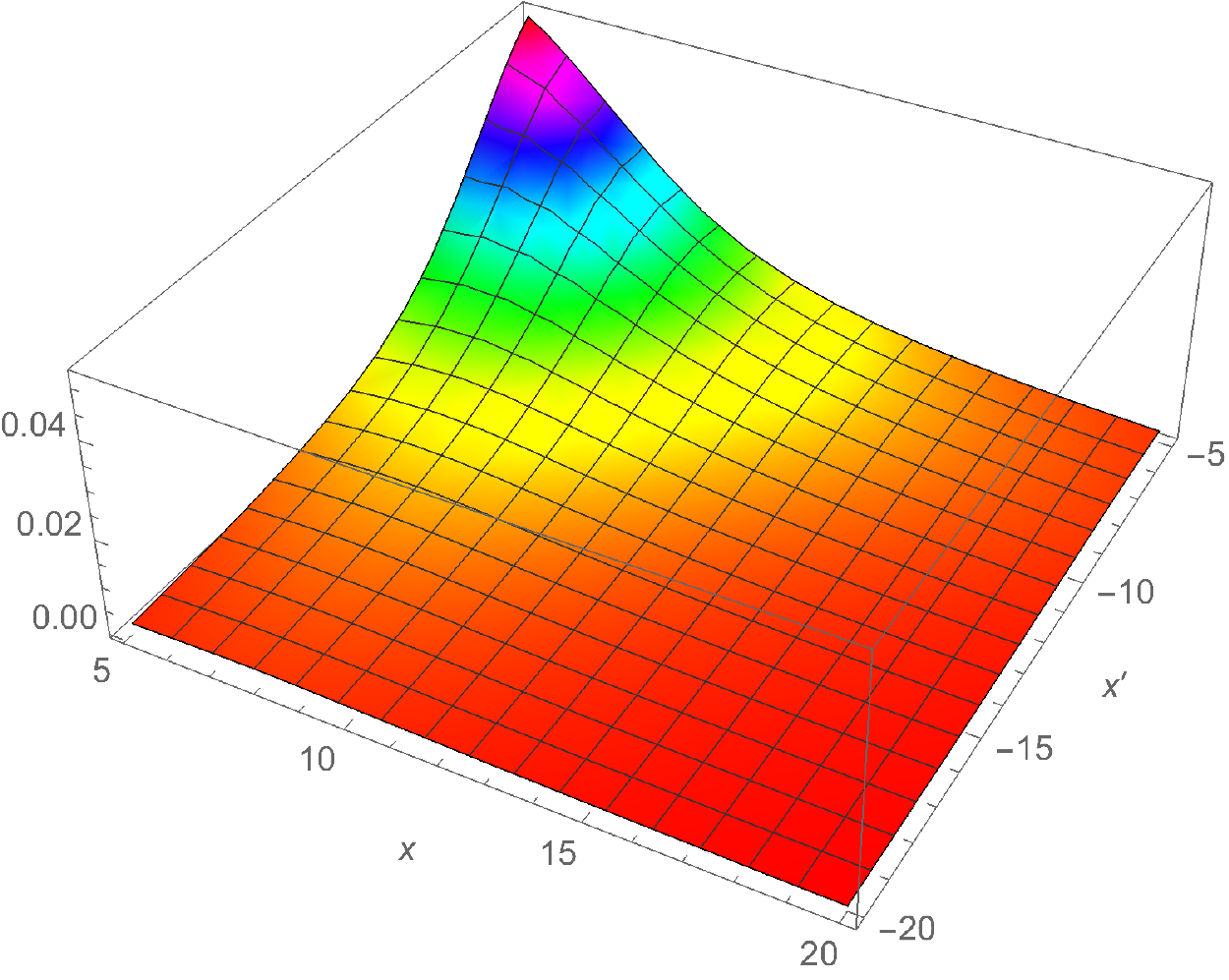}}
\caption{}
\label{6b}
\end{subfigure}
\begin{subfigure}[h]{0.45\textwidth}
{\includegraphics[width=2in]{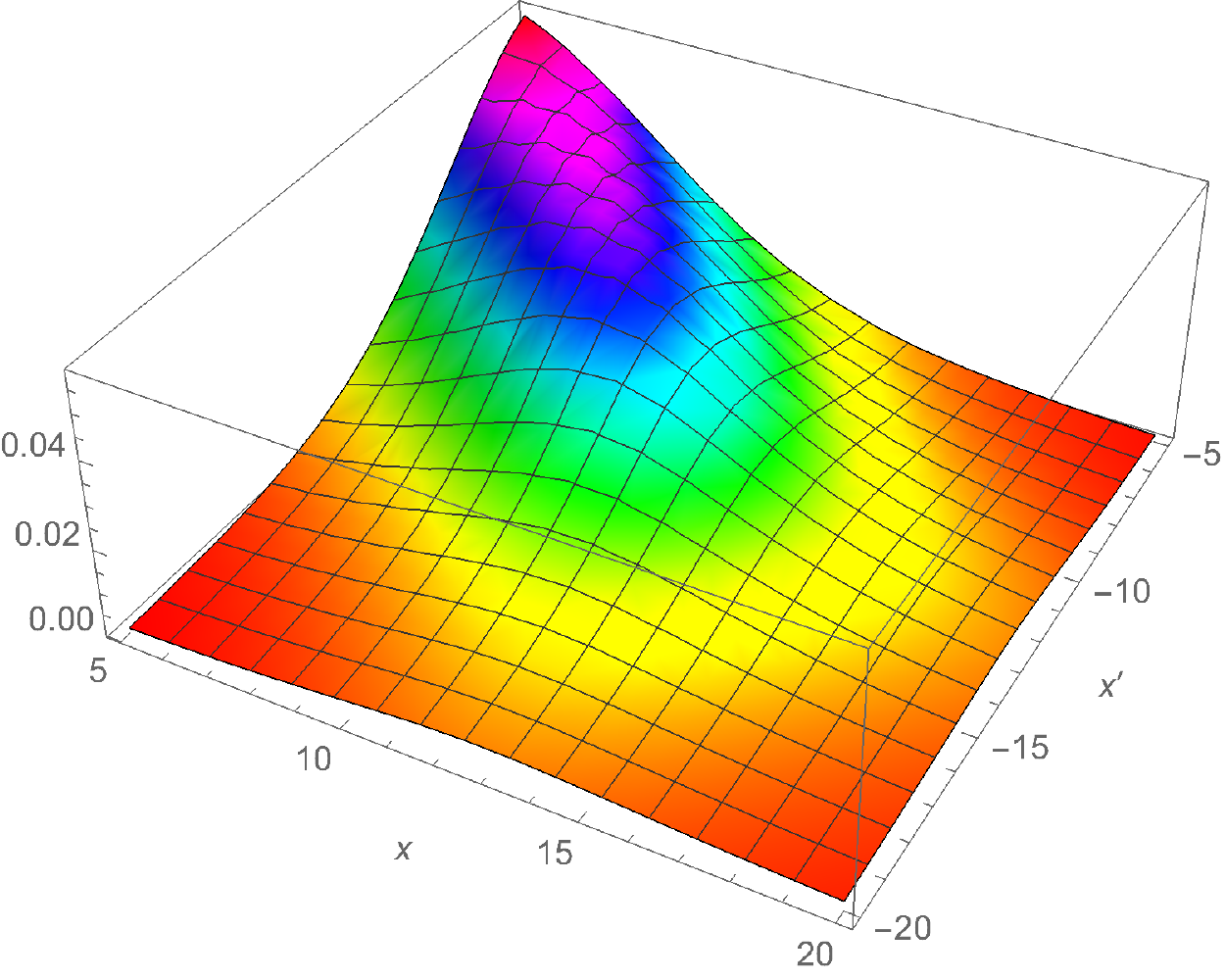}}
\caption{}
\label{6c}
\end{subfigure}
\begin{subfigure}[h]{0.45\textwidth}
{\includegraphics[width=2in]{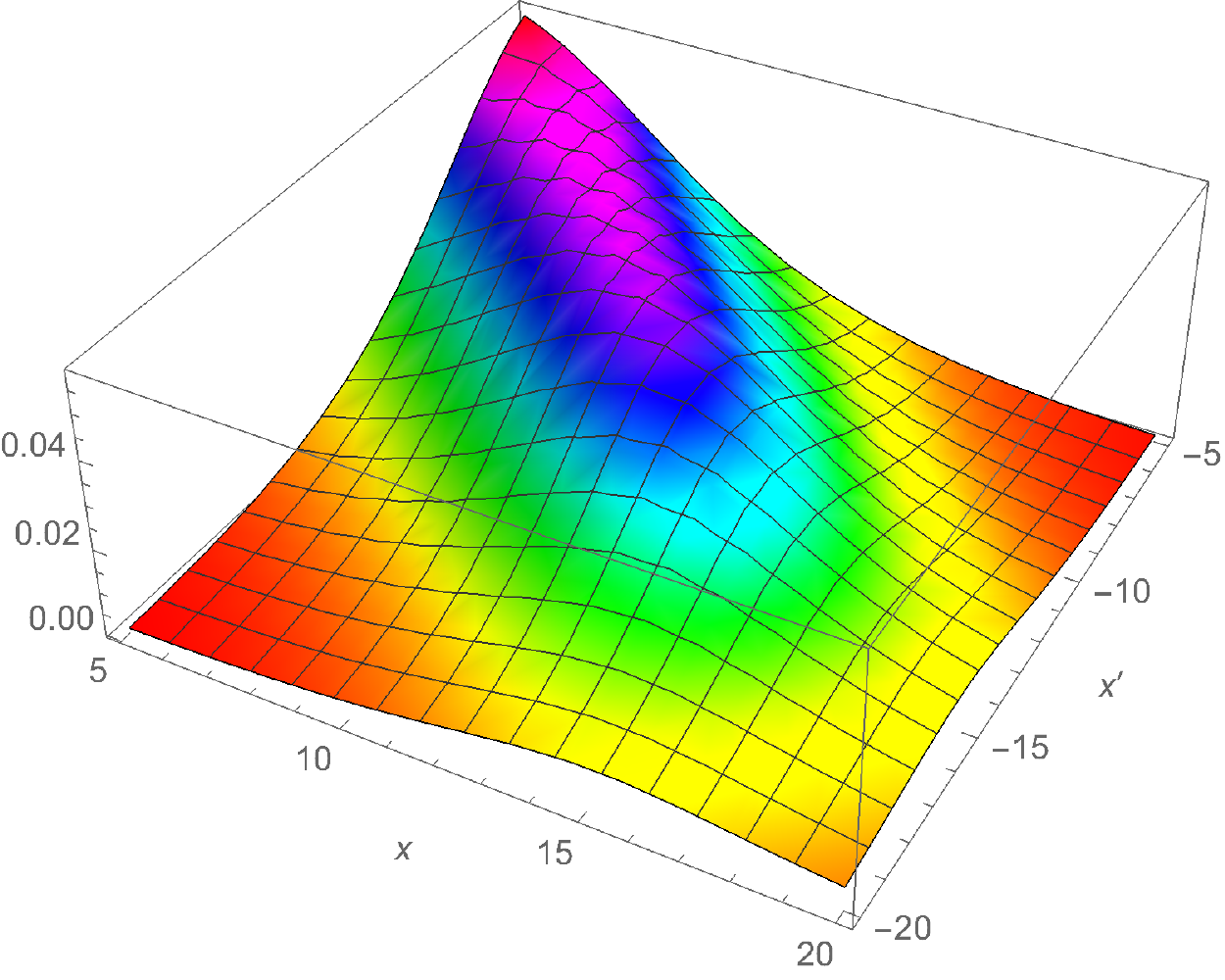}}\quad\quad
\caption{}
\label{6d}
\end{subfigure}
\caption{\label{fig6}3D plots of the absolute value of $G_{2\ T}^{(1)}(t;x,x')$ for $T=T_H$  at the same different \\ times of Fig. (\ref{fig1}), i.e. $t_1=1/\kappa$ (a), $t_2=2/\kappa$ (b), $t_3=4/\kappa$ (c), $t_4=5/\kappa$ (d).
 }
\end{figure}

\begin{figure}[h]
\centering
\begin{subfigure}[h]{0.45\textwidth}
{\includegraphics[width=2in]{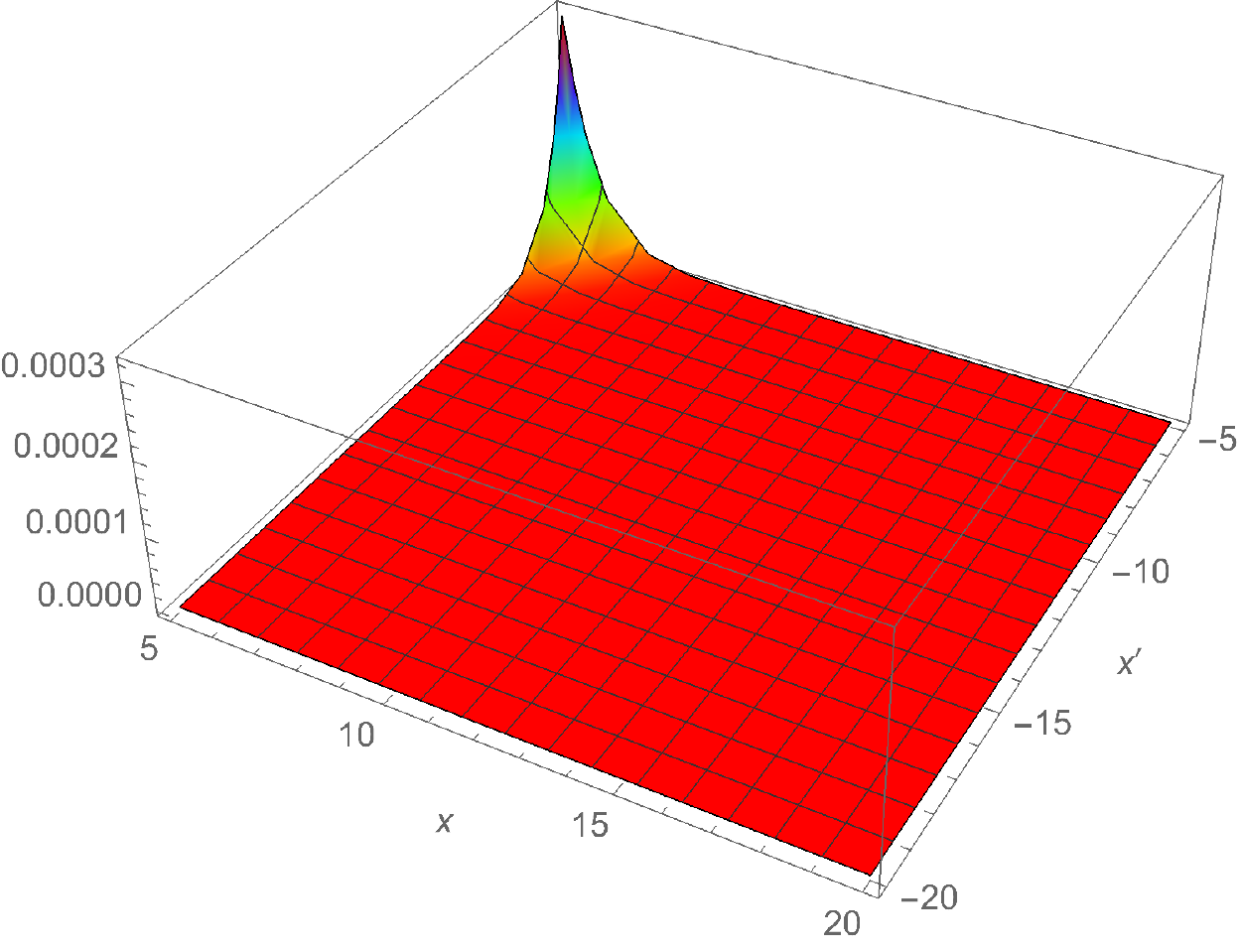}}
\caption{}
\label{7a}
\end{subfigure}
\begin{subfigure}[h]{0.45\textwidth}
{ \includegraphics[width=2in]{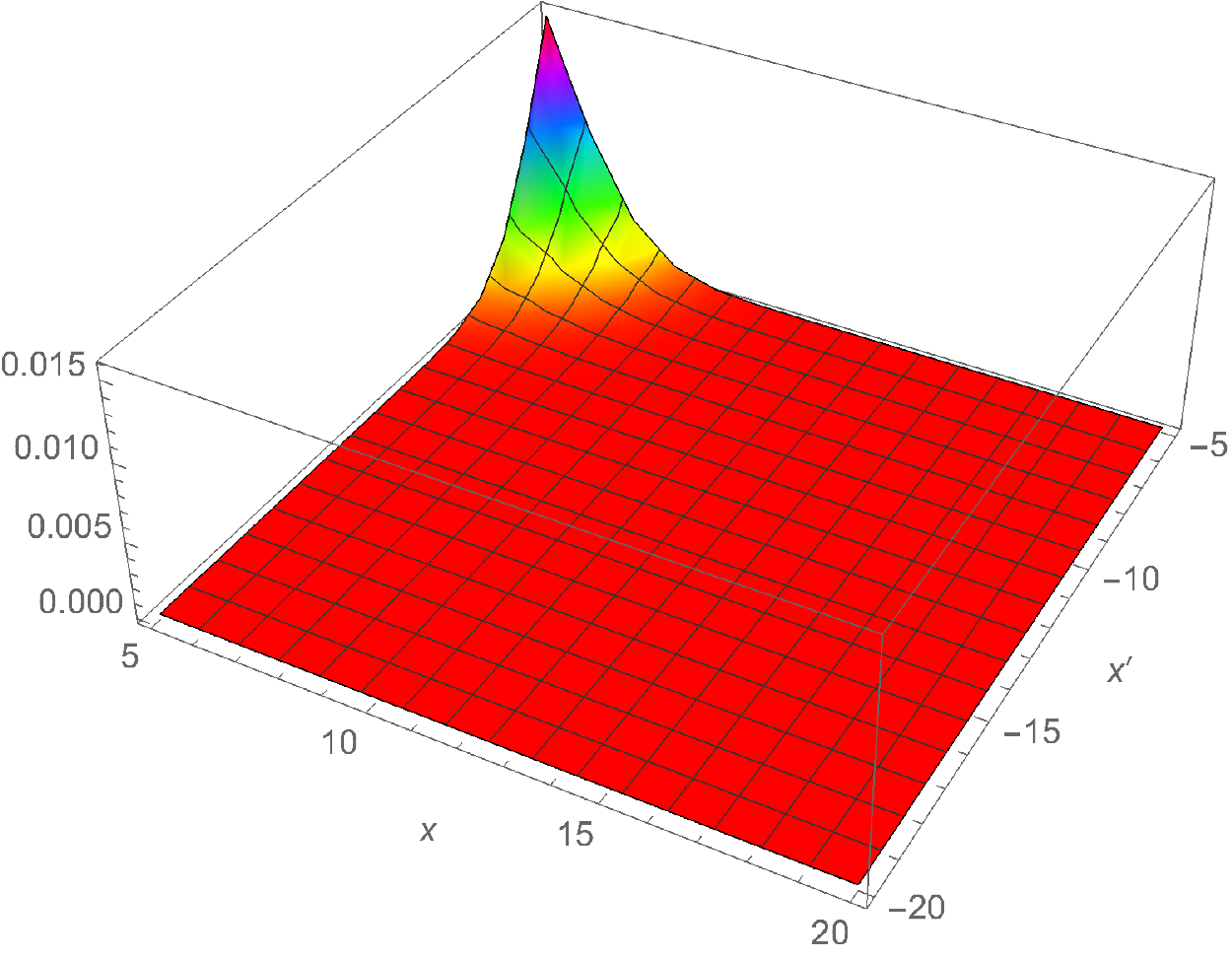}}
\caption{}
\label{7b}
\end{subfigure}
\begin{subfigure}[h]{0.45\textwidth}
{\includegraphics[width=2in]{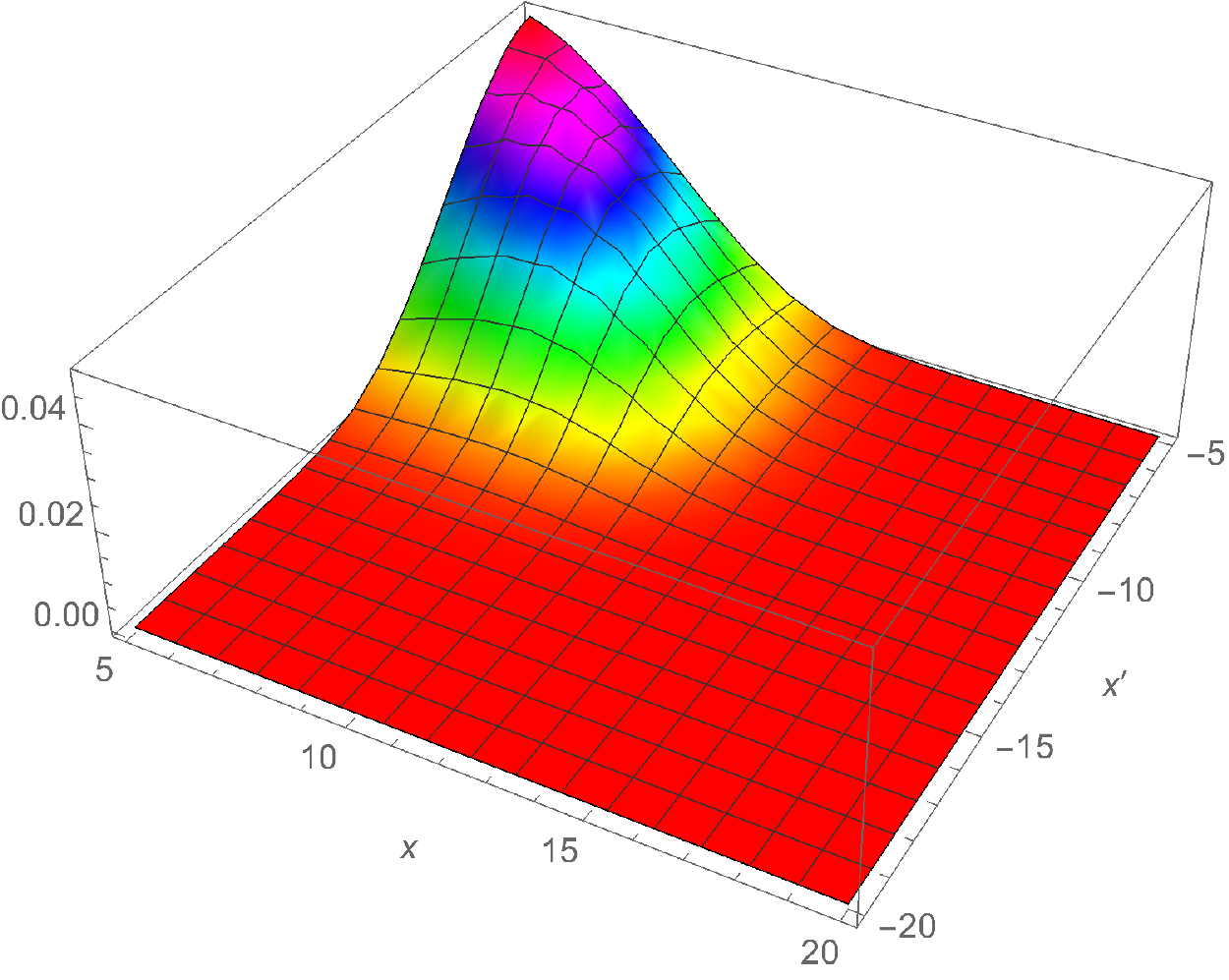}}
\caption{}
\label{7c}
\end{subfigure}
\begin{subfigure}[h]{0.45\textwidth}
{\includegraphics[width=2in]{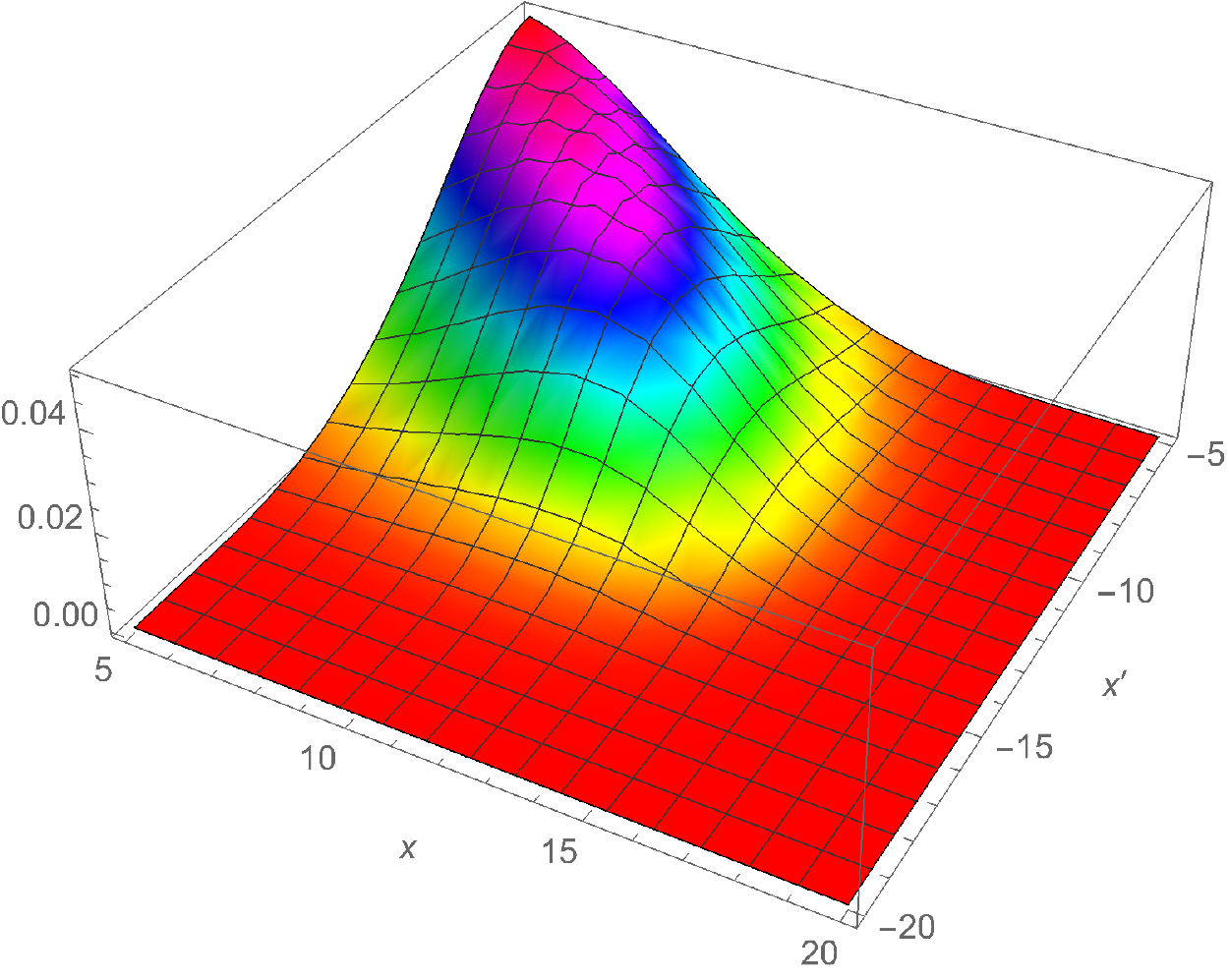}}\quad\quad
\caption{}
\label{7d}
\end{subfigure}
\caption{\label{fig7} 3D plots of the absolute value of $G_{2\ T}^{(1)}(t;x,x')$ for $T=10T_H$ at the same different\\  times of Fig. (\ref{fig1}), i.e. $t_1=1/\kappa$ (a), $t_2=2/\kappa$ (b), $t_3=4/\kappa$ (c), $t_4=5/\kappa$ (d).
    }
\end{figure}

In Fig. (\ref{fig6}), we have the three dimensional plots of $G_{2\ T}^{(1)}(t;x,x')$ for $T=T_H$ and in Fig. (\ref{fig7}) for $T=10T_H$ for the same times of Fig. (\ref{fig1}). 

\begin{figure}[h]
\centering
\begin{subfigure}[h]{0.45\textwidth}
{\includegraphics[width=3in]{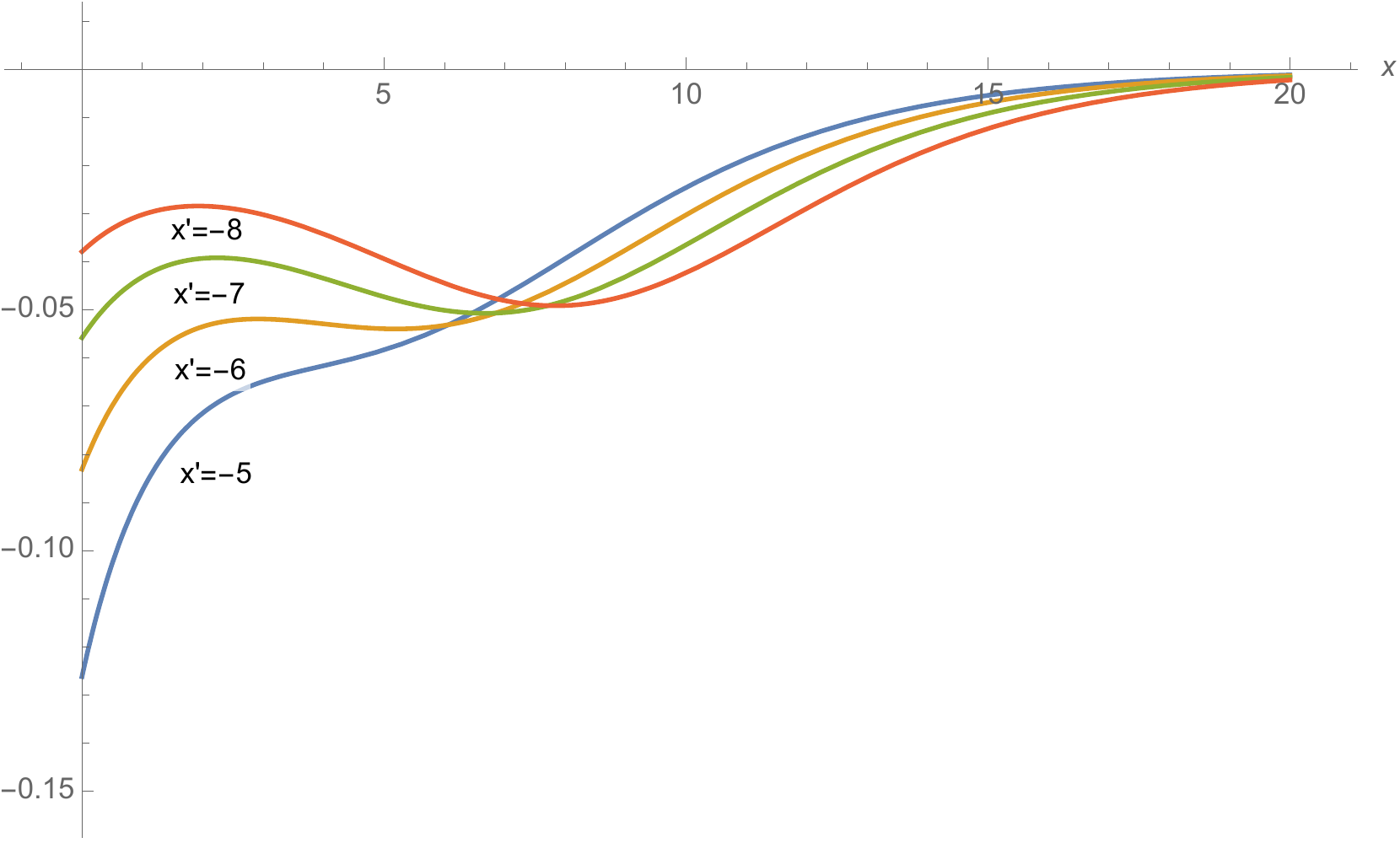}}
\caption{}
\label{8a}
\end{subfigure}\\
\begin{subfigure}[h]{0.45\textwidth}
{ \includegraphics[width=3in]{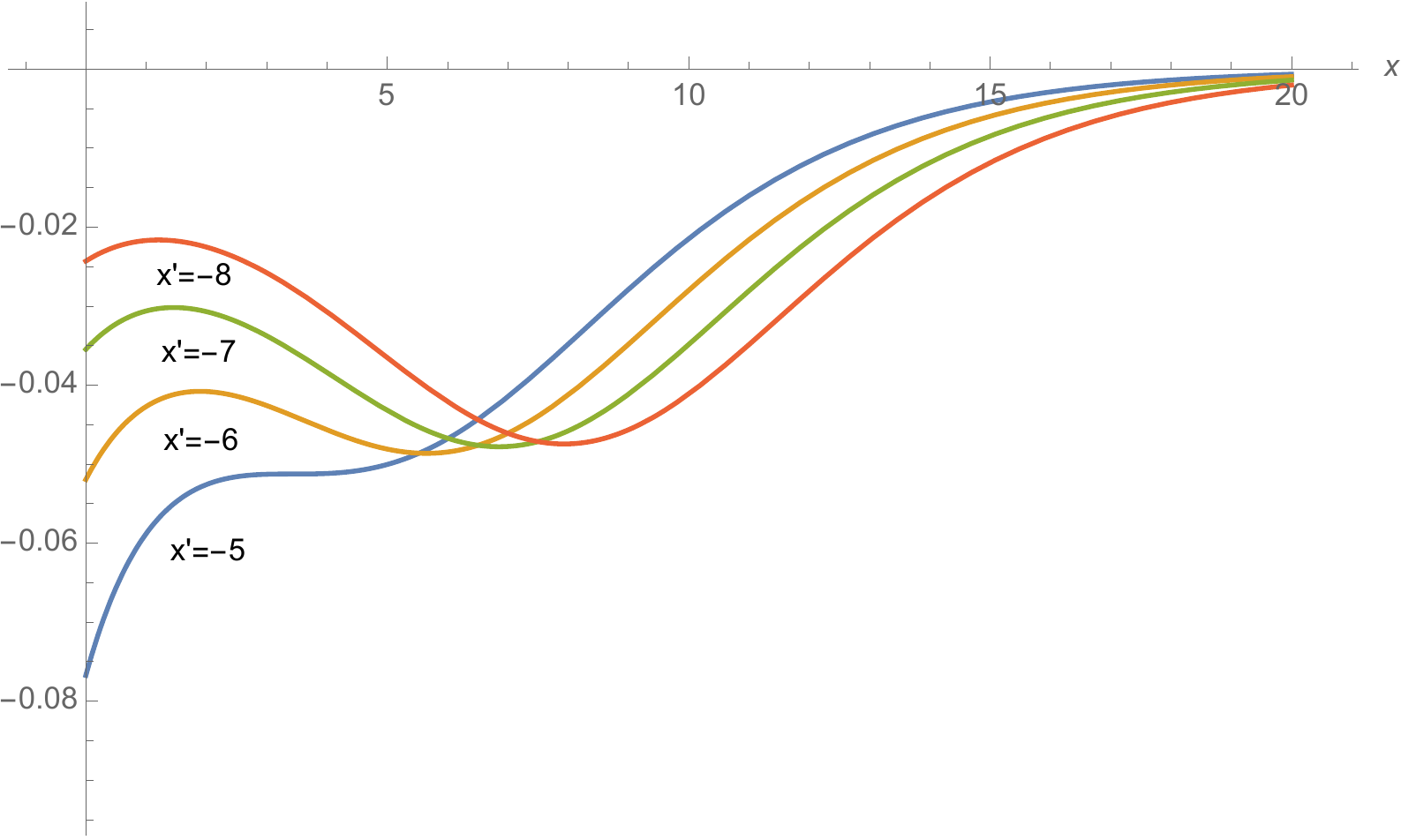}}
\caption{}
\label{8b}
\end{subfigure}
\caption{\label{fig8} Plots of $G_{2\ T}^{(1)}(t;x,x')$, $T=T_H$ in (a) and $T=10T_H$ in (b), at $t=\frac{10}{\kappa}$ as a\\ function of $x$ for various values of $x'$ as we did in Fig. (\ref{fig2}) for the $T=0$ case.   }
\end{figure}

We see that the ramp-up process to the stationary peaks configuration is slower. The delay grows as the temperature increases. For $T=10T_H$ the signal appears at $t\sim 5/\kappa$. 
In Fig. (\ref{fig8}) we show $G_{2\ T}^{(1)}(t;x,x')$, $T=T_H$ in Fig. (\ref{8a}) and $T=10T_H$ in Fig. (\ref{8b}), at $t=\frac{10}{\kappa}$ as a function of $x$ for various values of $x'$ as we did in Fig. (\ref{fig2}) for the $T=0$ case.

As it is shown in Appendix A (see eq. (\ref{atrentacinque})) at late time the first term in eq. (\ref{duedodici}), $G_{2\ T,u}^{(1)}(t;x,x')$, which is the dominant one giving the contribution of the $u$ sector, reduces to the corresponding one at $T=0$ in the same limit. This is a manifestation of the quantum version of the no-hair theorem for BHs (see for example \cite{cch}). Hawking radiation at late times is unaffected by any population in the initial state which causes just a transient stimulated emission \cite{wald76}. This because the modes responsable for the late time behaviour in the $u$ channel are the ones with $u\to \infty$, i.e. they propagate very close to the horizon and are highly redshifted in their journey towards the asymptotic region washing out any information on the initial state. The temperature dependence at late times comes only from the $v$ channel, i.e.  $G_{2\ T,v}^{(1)}(t;x,x')$, but its contribution to the total $G_{2\ T}^{(1)}(t;x,x')$ is negligible 
(for $T=T_H$ it is one order of magnitude smaller than $G_{2\ T,u}^{(1)}(t;x,x')$ and decreases by increasing $T$). Finite temperature corrections are therefore small within the gravitational approximation.  

\section{Gravitational BHs}

In this section, in analogy with what we did for the acoustic BEC BH, we shall investigate correlations between the Hawking particles and their partners across the horizon in a real gravitational BH. The striking difference is that the horizon is now a causal boundary preventing any information on inside events to leak outside. The study of these correlations has therefore only theoretical interest since one cannot perform experiments (even ``gedanken'' ones) which require measurements whose results can be exchanged by the observers both in the exterior and in the interior of a BH. Nevertheless, as we shall see, the theoretical results of our investigation will turn out to be, at a preliminary inspection,  rather unexpected and in strong disagreement with what one finds in the BEC case reviewed in the previous section.   

However one should keep in mind that, although the near horizon geometry of an acoustic BH is similar to the real one of a gravitational BH, the inner region of an acoustic BH is in principle (at least in the experiments so far performed) infinite, while the one of a real BH terminates rapidly in a singularity. This fact has a tremendous impact on the correlation functions since these are genuinely nonlocal quantum objects and as such they do not probe just the local spacetime geometry. 

\begin{figure}[h]
\includegraphics[width=6.5in]{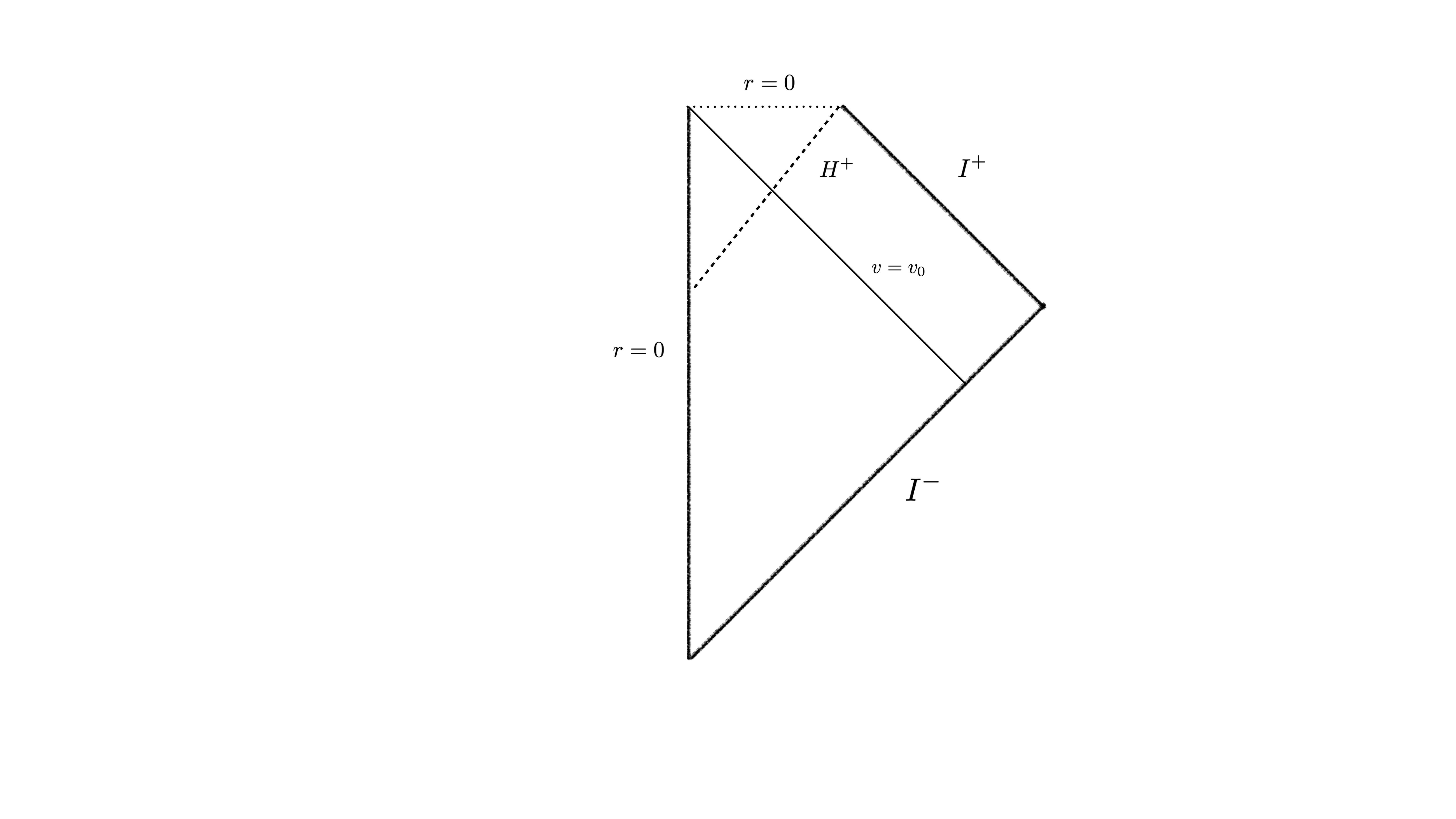} 
\caption{\label{fig9}  Penrose diagram of a Schwarzschild black hole formed by a null shell collapse.}
\end{figure}

We shall begin by considering a very simple toy model of BH formation widely used in the gravitational literature, namely the collapse of a null shell of radiation (see for example \cite{libro-2005}). The shell is located at $v=v_0$, where $v$ is an advanced null coordinate. The space-time metric of this model can be given simply in the Vaidya form
\be ds^2=-(1-\frac{2M(v)}{r})dv^2+2dvdr+r^2(d\theta^2+\sin^2\theta d\varphi^2)\ , \label{treuno} \ee
where the mass function  $M(v)$  has this form
\be M(v)=m\theta(v-v_0)\ , \label{tredue} \ee
where $m$ is a constant and $\theta$ the Heaviside step function. So inside the shell (i.e. $v<v_0$) we have Minkowski space-time
\be ds^2_{in}=-dv^2+2dvdr+r^2d\Omega^2\ , \label{tretre} \ee
where $d\Omega^2=d\theta^2+\sin^2\theta d\varphi^2$, while outside the shell (i.e. $v>v_0$) the metric is the Schwarzschild one describing a BH of mass $m$ and horizon located at $r=2m$
\be ds^2=-(1-\frac{2m}{r})dv^2+2dvdr+r^2d\Omega^2\ . \label{trequattro} \ee
Continuity of the induced metric on the shell ensures that the radial coordinate $r$ is continuous across the shell. The resulting Penrose diagram of our spacetime is given in Fig.  (\ref{fig9}).

In the interior region we can introduce a retarded null coordinate $u_{in}$ as
\be u_{in}=v-2r \label{trecinque} \ee
and write the Minkowski metric in a double null form 
\be 
ds^2=-du_{in}dv+r^2(u_{in},v)d\Omega^2 \label{tresei}\ ,
\ee
where 
\be r(u_{in},v)=\frac{v-u_{in}}{2}\ . \label{tresette}
\ee
Similarly in the external region we define the retarded Eddington-Finkelstein null coordinate $u$ as
\be u=v-2r^* \label{treotto} \ee
where $r^*$ is the tortoise Regge-Wheeler coordinate
\be r^*=\int \frac{dr}{1-\frac{2m}{r}}=r+2m\ln|\frac{r}{2m}-1| \label{trenove} \ee
and the double null form of th Schwarzschild metric reads
\be ds^2=-(1-\frac{2m}{r})dudv+r^2(u,v)d\Omega^2 \ , \label{tredieci} \ee
where $r(u,v)$ is implicitly defined by 
\be r^*=\frac{v-u}{2}\ . \label{treundici} \ee
Along the shell we have $r(u_{in},v_0)=r(u,v_0)$ and this leads to 
\be u=u_{in}-4m\ln|\frac{v_0-4m-u_{in}}{4m}|\ . \label{tredodici} \ee
For simplicity we set $v_0=4m$. This relation, once inverted as $u_{in}(u)$, allows to extend the coordinate $u_{in}$ in the exterior region. In particular we have \cite{gae}
\be u_{in}=-4mW(\pm e^{-\frac{u}{4m}})\ , \label{tretredici} \ee
where $W$ is the Lambert function and $+$ holds in the region exterior to the horizon and $-$ in the interior one. 

We consider now a quantized field $\frac{\hat\phi}{\sqrt{4\pi}r}$, for simplicity scalar and massless, propagating in this spacetime. The modes associated to this field are assumed to have the form $\frac{e^{-i\omega v}}{\sqrt{4\pi}\sqrt{2\pi\omega}r}$ on past null infinity $I^-$, in this way the quantum state of our field is Minkowski vacuum on $I^-$, i.e. there are no incoming particles. We call this state $|in\rangle$. We neglect backscattering of the modes induced by the curvature of the spacetime and simply impose reflecting conditions on the origin $r=0$ in the Minkowski region yielding $u_{in}=v$ at $r=0$ (see eq. (\ref{tresette})) and requiring regularity of the modes there. This leads to approxinate our (1+3) dimensional theory as an effective (1+1) dimensional one describing a massless scalar field $\hat\phi$ propagating in the (1+1) dimensional spacetime section of our spacetime obtained by taking $\theta$ and $\varphi$ constants, i.e. 
\be ds^2_{(2)}=-(1-\frac{2M(v)}{r})dv^2+2dvdr\ . \label{trequattordici} \ee
This procedure is widely used in dealing with the Hawking effect for the Schwarzschild BH allowing an analytical description which captures the essential physical features of the process. See for instance  \cite{dfu, bb, pp}.

The two-point function corresponding to our $|in\rangle$ vacuum accounting for the boundary condition at $r=0$ in the Minkowski region reads \cite{louko}
\be
\langle in| \hat \phi(x)\hat \phi (x') |in\rangle =-\frac{\hbar}{4\pi}\ln \frac{(u_{in}-u_{in}')(v-v')}{(u_{in}-v')(u_{in}'-v)}\ . \label{trequindici} \ee
In this section we indicate with $x$ the generic space-time point $(t,r)$. 
Unlike the previous case of an acoustic BH, spacetime is now really curved and this is affecting not just the quantum field $\hat\phi$, but also the observers which are requested to probe it. So we have to specify not only the observable of the field $\hat\phi$ we want to measure but also which observer is going to measure it. 

In the acoustic BH one considers the density fluctuations measured by an observer at rest in the laboratory, which is an inertial observer in that case. To be as close as possible, we can choose the energy density of the field $\hat \phi$ as quantum observable but one cannot choose observers at rest, since they do not exist inside the horizon and we are interested in correlations across the horizon which require measurements on both sides of it.

We therefore choose free falling observers, moving on radial geodesics for simplicity. These are local inertial observers. So our quantum observable is the scalar
\be \hat \rho\equiv \hat T_{ab}u^au^b \ , \label{tresedici} \ee
where 
\be \hat T_{ab}(\hat\phi)=\partial_a\hat \phi\partial_b\hat \phi-\frac{g_{ab}}{2}g^{cd}\partial_c\hat \phi\partial_d\hat \phi \label{trediciassette} \ee
is the two-dimensional energy momentum operator of the field $\hat\phi$ and $u^a$ is the four velocity of the free falling observer. Note that $\hat T_{ab}$ is traceless due to the conformal invariance of $\hat\phi$  so it has only two independent components. 
The correlator we shall study is then
\be \langle in|\hat \rho(x)\hat\rho (x')|in\rangle\equiv  G(x,x') \label{trediciotto} \ee
evaluated in the Schwarzschild region where one point ($x$) is taken outside the horizon and the other ($x'$) inside. 

Since in the acoustic BH the effective metric is given in Painlev\'e coordinates, we use the same coordinates here transforming the original Schwarzschild metric (\ref{trequattro}) into
\be ds^2_{(2)}=-fdt^2-2Vdtdr+dr^2\ , \label{trediciannove} \ee
where $r=1-\frac{2m}{r}\equiv (1-V^2),\ V=-\sqrt{\frac{2m}{r}}$, and the Painlev\'e time $t$ is given by
\be t=v+\int (\frac{\sqrt{1-f}}{f}-\frac{1}{f})dr\ . \label{trediciannovebis} \ee
The metric (\ref{trediciannove}) holds for points ($t,r$) such that $v(t,r)>v_0$, where $v(t,r)$ is obtained inverting (\ref{trediciannovebis}). $t$ is a regular time coordinate across the horizon. With this choice we can slice the entire BH spacetime by constant $t$ hypersurfaces similarly with what we did for the acoustic BH.\footnote{Instead of $t$ we can use as time parameter Eddington-Finkelstein time $t_{EF}=v-r$ which is also a regular time coordinate across the horizon and slice the spacetime accordingly. No qualitative difference emerges in the results we will obtain.}

As shown in Appendix (\ref{appendixB}) the energy density measured by a free falling observer in Painlev\'e coordinates is simply
\be \hat\rho(t,r)=\hat T_{rr}(\hat\phi)\ . \label{treventi} \ee
Using eq. (\ref{bsette})  we can write the density-density correlator as
\bea G(x,x') &=& \langle in| \frac{\hat T_{uu}(x)\hat T_{u'u'}(x')}{(1+V(x))^2(1+V(x'))^2}+
\frac{\hat T_{uu}(x)\hat T_{v'v'}(x')}{(1+V(x))^2(1-V(x'))^2}\nonumber \\
&+& \frac{\hat T_{vv}(x)\hat T_{u'u'}(x')}{(1-V(x))^2(1+V(x'))^2}+\frac{\hat T_{vv}(x)\hat T_{v'v'}(x')}{(1-V(x))^2(1-V(x'))^2} |in\rangle\ .\label{treventuno} \eea

\begin{figure}[h]
\centering
\begin{subfigure}[h]{0.45\textwidth}
{\includegraphics[width=2in]{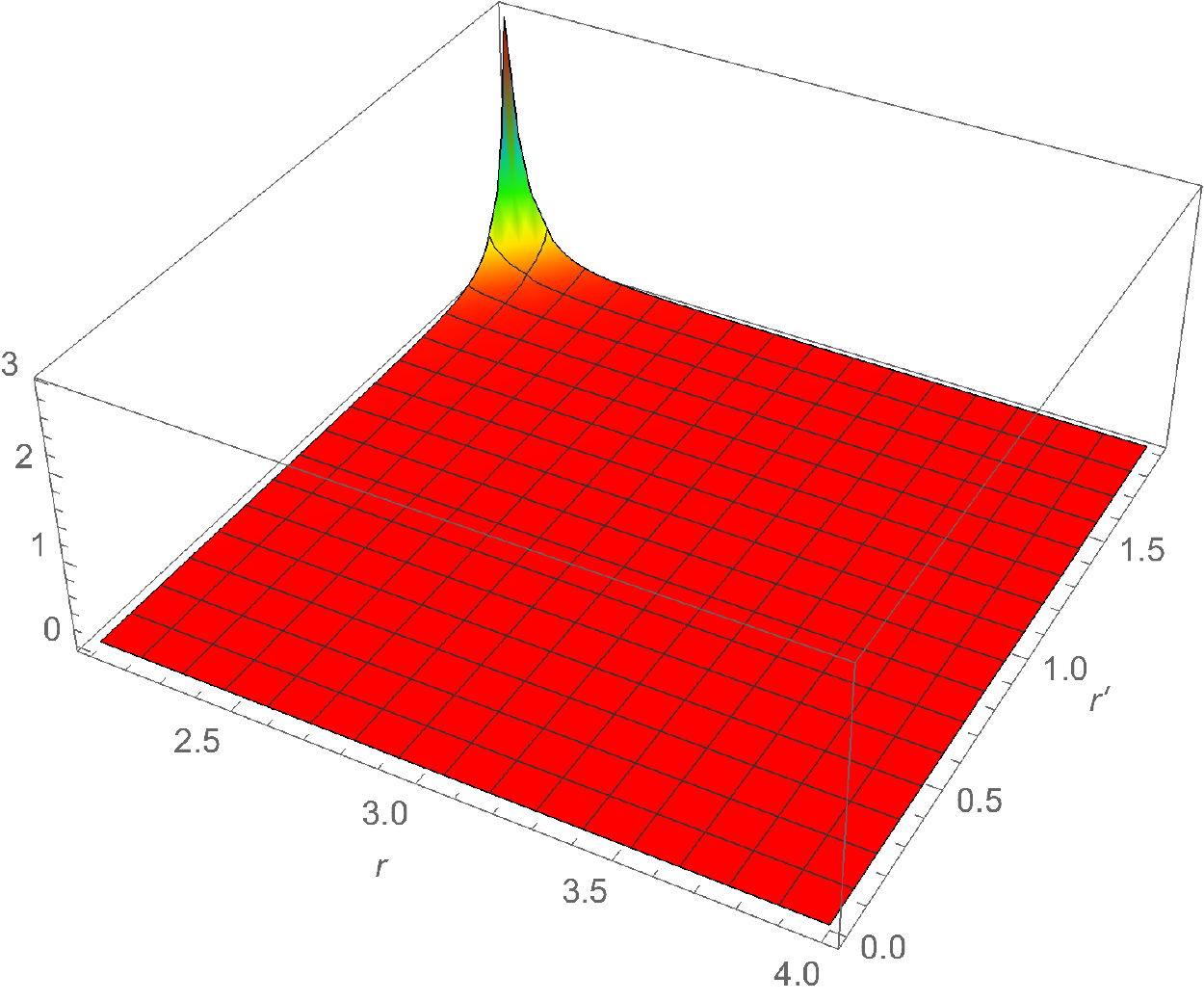}}
\caption{}
\label{10a}
\end{subfigure}
\begin{subfigure}[h]{0.45\textwidth}
{ \includegraphics[width=2in]{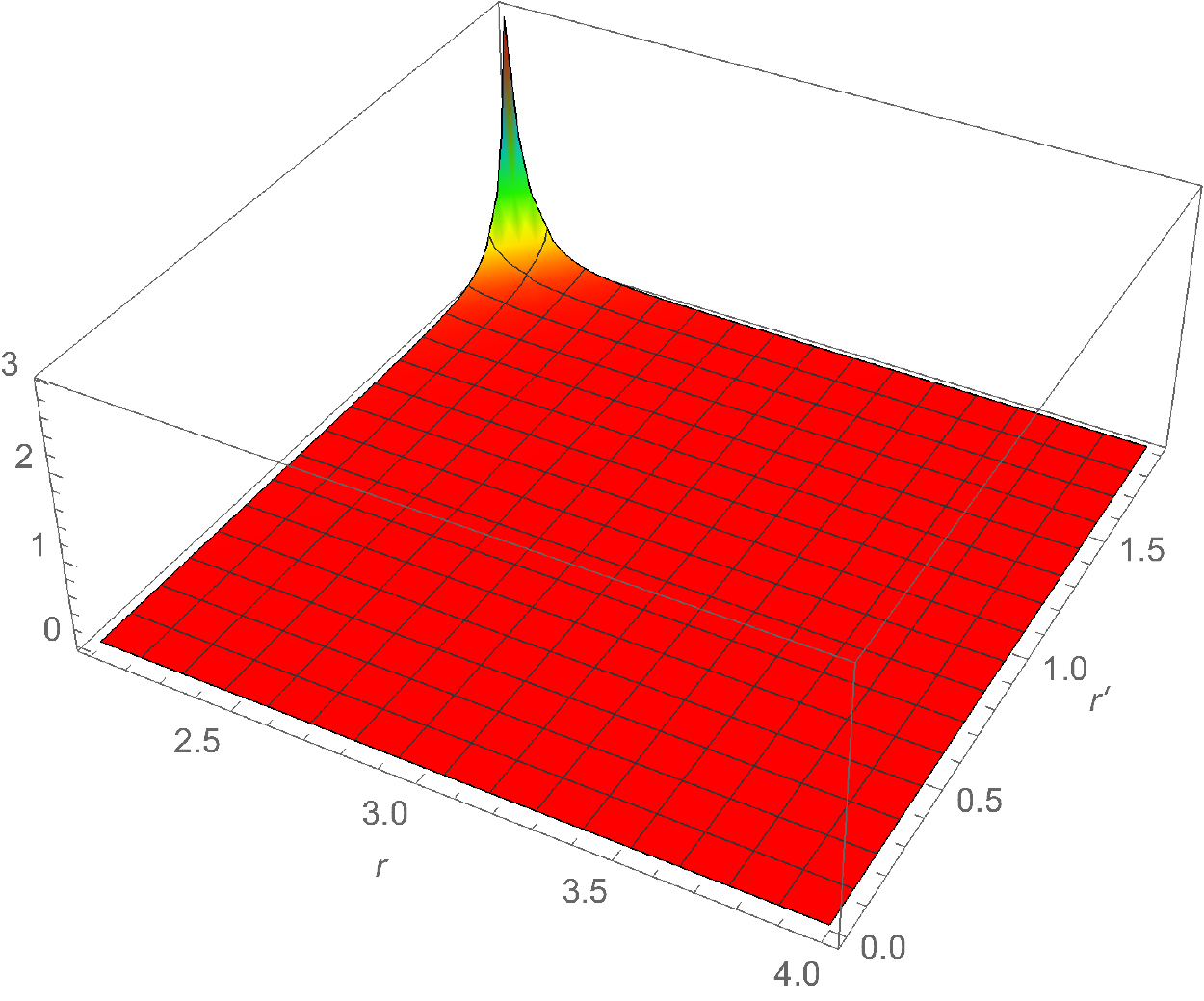}}
\caption{}
\label{10b}
\end{subfigure}
\begin{subfigure}[h]{0.45\textwidth}
{\includegraphics[width=2in]{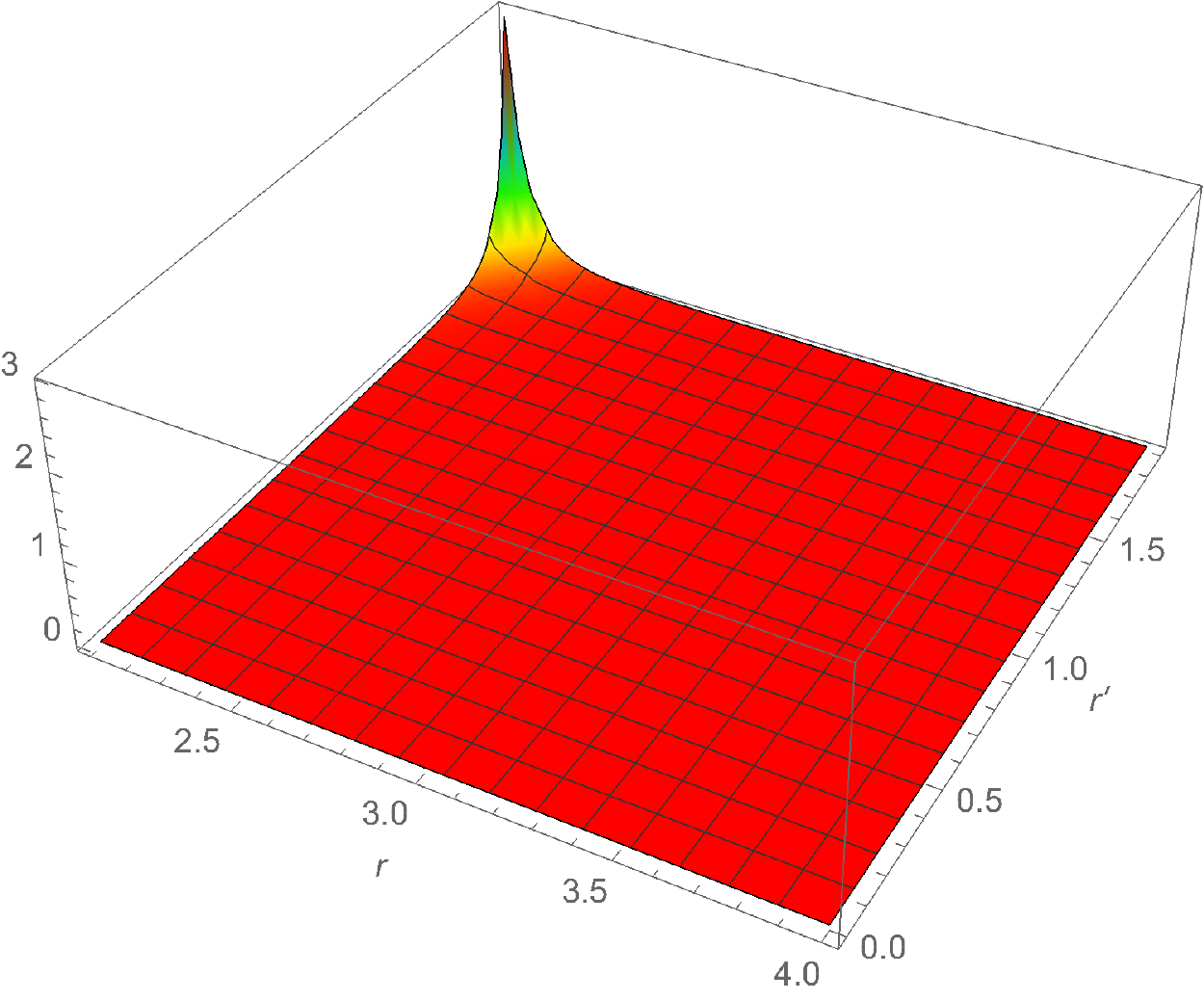}}
\caption{}
\label{10c}
\end{subfigure}
\begin{subfigure}[h]{0.45\textwidth}
{\includegraphics[width=2in]{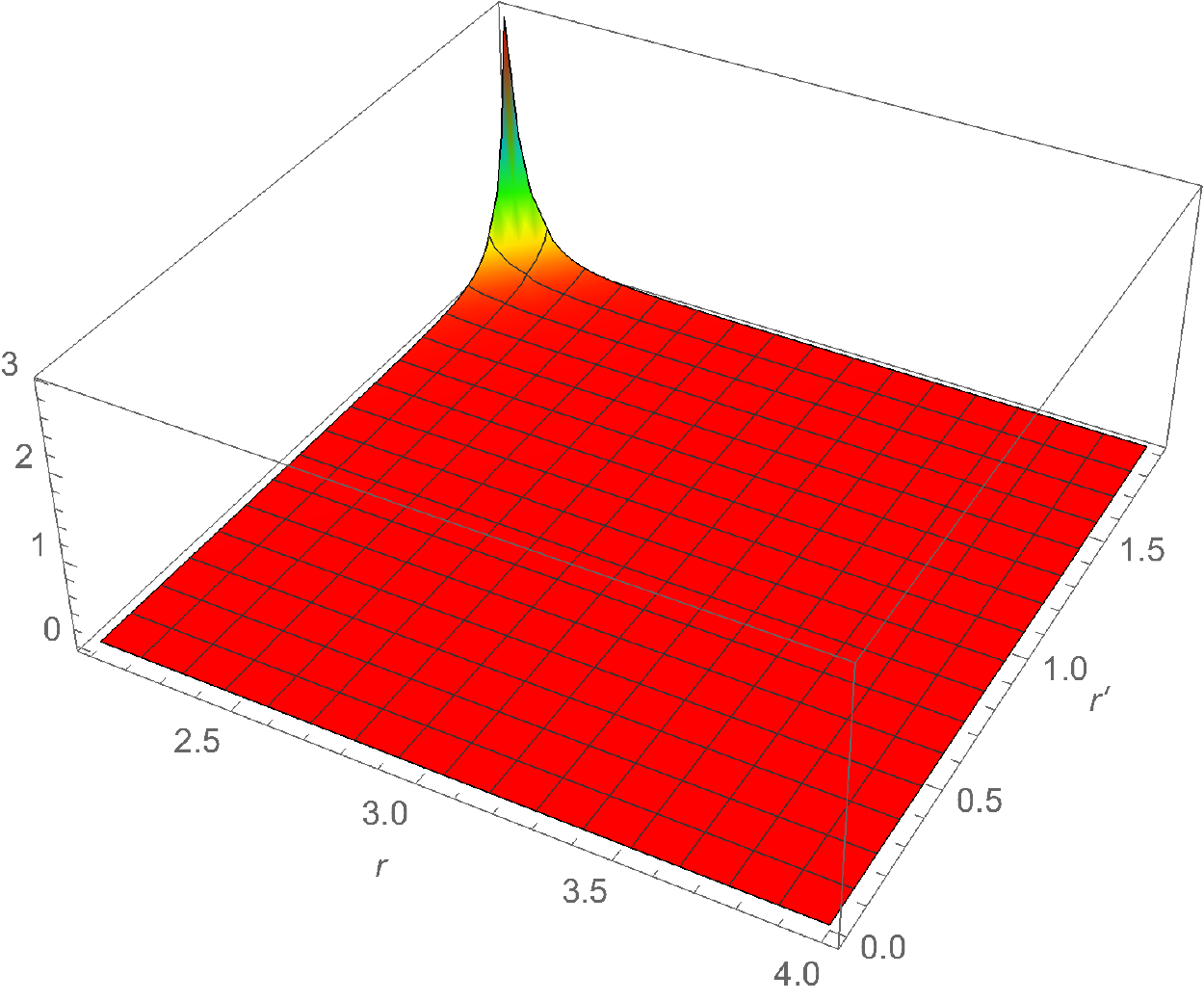}}\quad\quad
\caption{}
\label{10d}
\end{subfigure}
\caption{\label{fig10} Plot of $G(t,r;t',r')$ ($\hbar=1, m=1$ in all plots of this section)  at\\ equal Painlev\'e time for $0<r'<2m$ and $2m<r<4m$ outside the shell\\ at four increasing Painlev\'e times $t=10m$ (a), $20m$ (b), $30m$ (c), $40m$ (d).  }
\end{figure}

In Fig. (\ref{fig10}) we plot the correlator $G(t,r;t',r')$ at equal Painlev\'e time (i.e. $t=t'$) for $0<r'<2m$ and $2m<r<4m$ outside the shell. The sequence of figures corresponds to four increasing Painlev\'e times $t=10m,\ 20m,\ 30m,\ 40m$. 

\begin{figure}[h]
\includegraphics[width=3in]{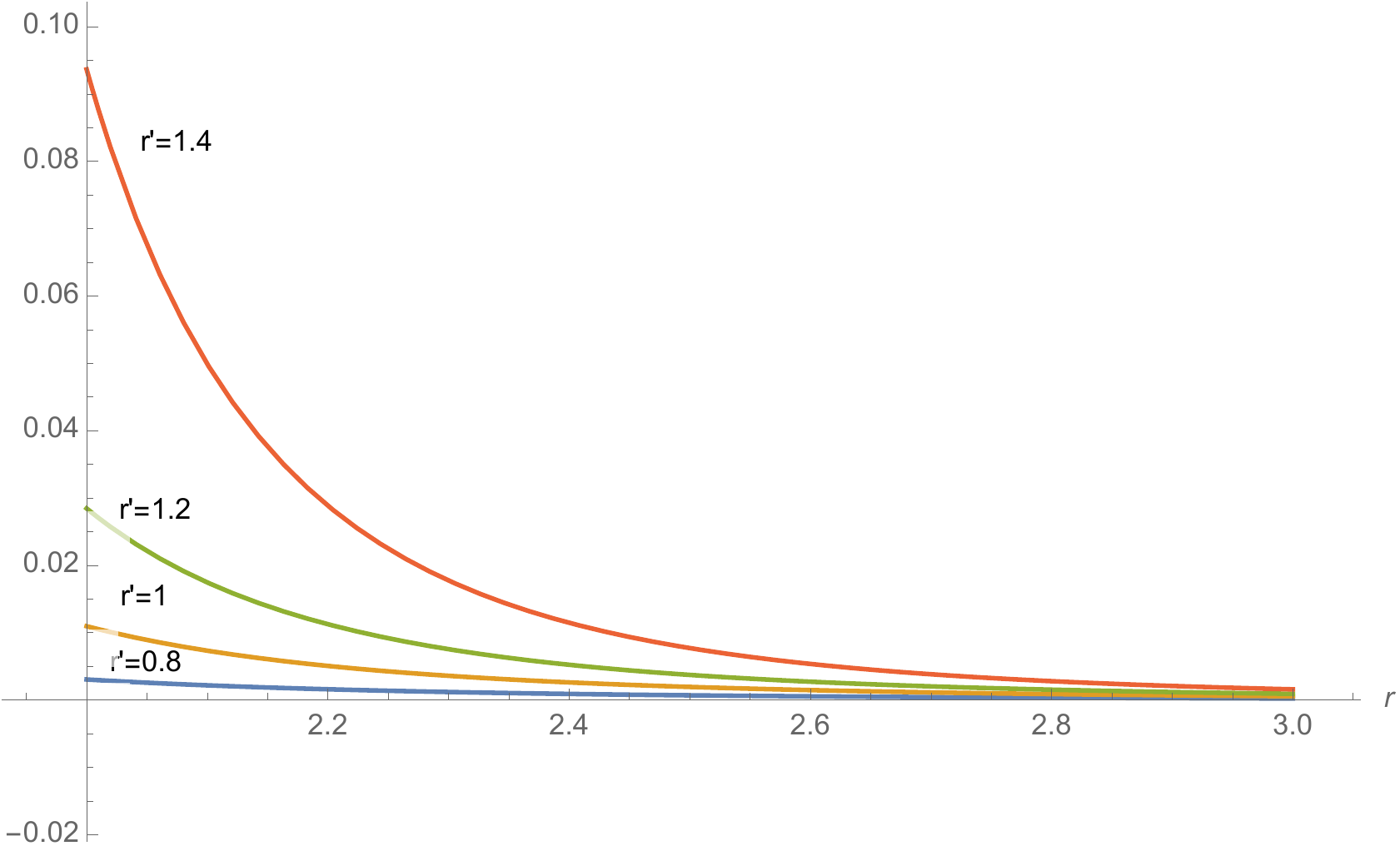} 
\caption{\label{fig11}  $G(t,r;t',r')$ at $t=t'=40m$ as a function of $r$ at values of $r'<2m$: $r'=0.8m, m,  1.2m, 1.4m$.  }
\end{figure}

The result is quite surprising: no structure at all appears.\footnote{Although all Figs. (\ref{fig10}) all seem the same, differences 
nevertheless exist away from the divergent coincident limit $r=r'=2m$, but they are too 
small to be seen in these plots and are irrelevant for our discussion.} The same conclusion is reached by representing the correlator at $t=t'=40m$ as a function of $r$ for different values of $r'<2m$: $r'=0.8m,\ m,\ 1.2m,\ 1.4m$. See Fig. (\ref{fig11}).

So it seems that there is no signal in then equal time correlator (3.22) of the correlations between the Hawking particles and their partners, and this in striking contradiction with our naive expectation and with what we found for the acoustic BH. A comparison of Figs. (\ref{fig1},\ref{fig2}) of our previous section with the actual ones is clearly illustrating this discrepancy. For the acoustic BH we found no sign of the correlations close to the horizon because there the correlator is dominated by the light cone (coincidence limit in the case of equal time) singularity. Correlations appeared however well outside the horizon with the characteristic peak we saw. So our expectation was to find a similar peak structure even in our gravitational BH for points sufficiently away from the horizon, in the ``quantum atmosphere" region. 

\begin{figure}[h]
\includegraphics[width=3in]{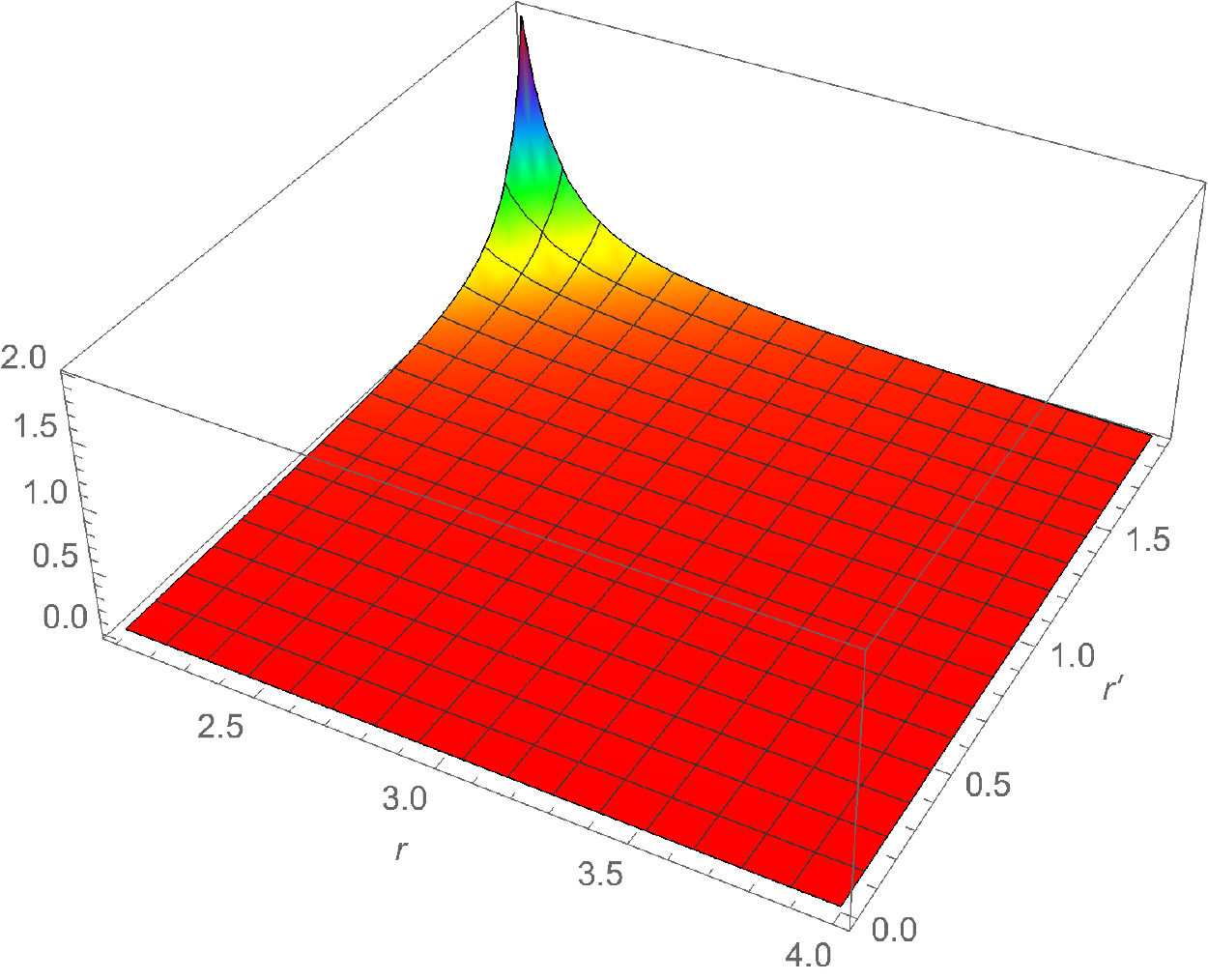} 
\caption{\label{fig12}  Density correlator (\ref{duesei}) for an hypothetical acoustic Schwarzschild metric.  }
\end{figure}

So one has to understand why this feature does not show up.
At first sight, one could argue that this is an artifact related to the motion of the observers (free falling in our case) measuring the correlations affecting the result, while in the acoustic case measurements are made at fixed laboratory coordinates. We do not think this is the case. If one calculates the equal time density-density correlator of eq. (\ref{duesei})  for an hypothetical acoustic metric given by the Schwarzschild one of eq. (\ref{trequattro}) for which $c=1$ and the velocity profile is
$V=-\sqrt{2m/r}$ (diverging at $r=0$) and where the $r=0$ singularity is replaced by a sink  absorbing both the condensate atoms and the phonons, as discussed for example in  Ref. \cite{cond-sch}, one would obtain the plot depicted in Fig. (\ref{fig12}). The same negative result: no peak in the equal time correlation function.

Now the key ingredient for the description of these correlations  (see eqs. (\ref{treventuno}), (\ref{bdieci},  (\ref{bquattordici})) is the following correlator
\be \label{tre25} \frac{ \partial_u\partial_{u'} \langle in|\hat\phi(x)\hat\phi(x')|in\rangle }{(1+V(r))(1+V(r'))}|_{t=t'}=-\frac{\hbar}{4\pi   (1+V(r))(1+V(r'))}
\frac{u_{in}u_{in}'}{(u_{in}-u_{in}')^2} \frac{1}{(u_{in}-4m)(u_{in}'-4m)}|_{t=t'} \ee
from which the relevant part of the density-density correlations (eq. (\ref{treventuno})) is constructed.
Note its similarity with the $G_2^{(1)}(t;x,x')$ 
correlator of eq. (\ref{duesei}) in the acoustic case discussed in the previous section.

\begin{figure}[h]
\includegraphics[width=3in]{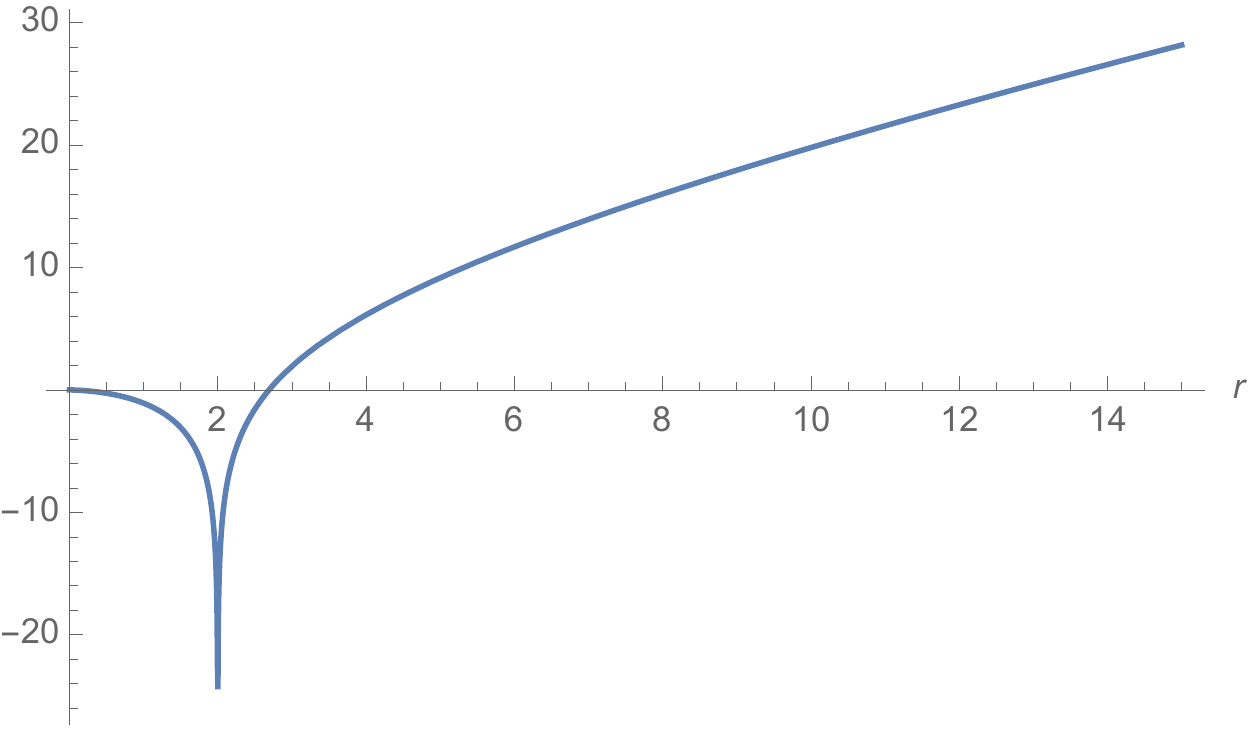} 
\caption{\label{fig13}  Plot of the left and right hand side of eq. (\ref{tre29})  }
\end{figure}

\begin{figure}[h]
\includegraphics[width=3in]{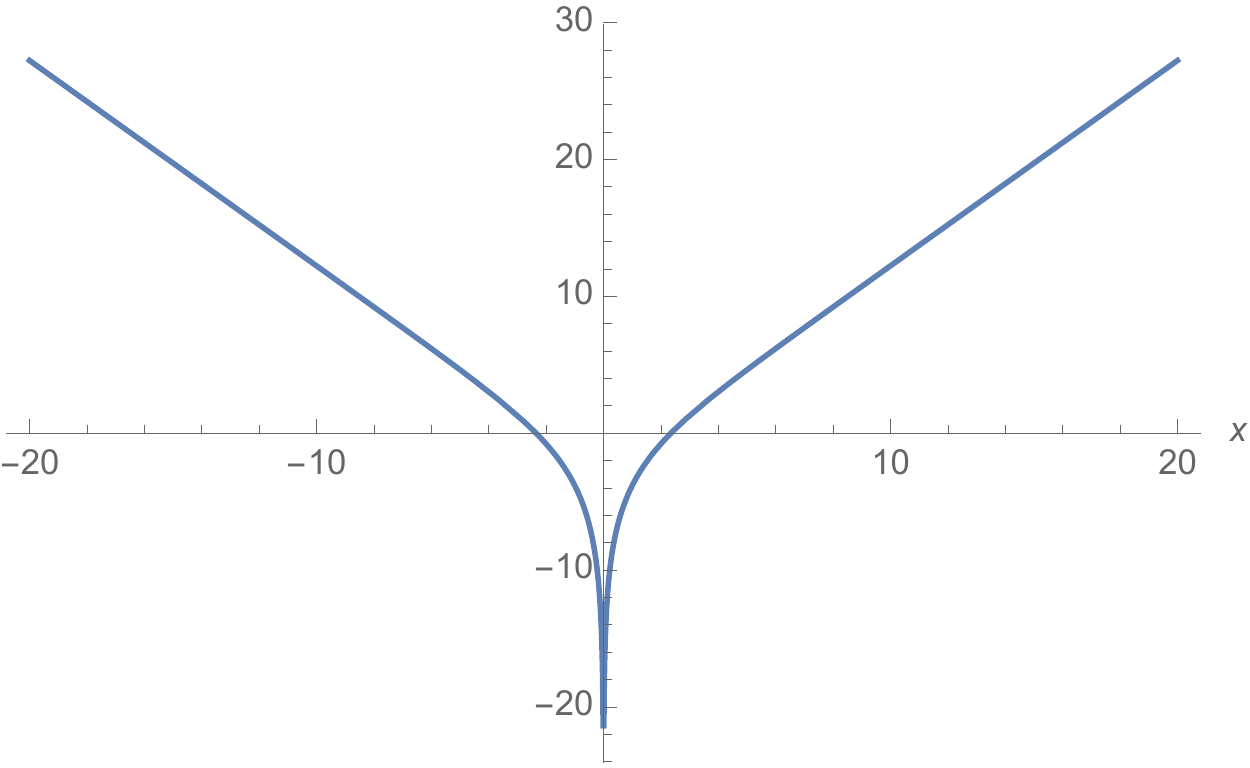} 
\caption{\label{fig14}  Plot of the $u=u'$ condition at $t=t'$ in the acoustic case, with $u$ given in (\ref{adieci}). }
\end{figure}

 At late retarded time ($u\to +\infty, u_{in}\to 0$) we have, from (\ref{tredodici}), that the relation between $u_{in}$ and $u$ can be approximated as
\be\label{tre26} u_{in}=\pm 4m e^{-\frac{u}{4m}} \ , \ee
where $u_{in}<0$ for $r>2m$ and $u_{in}>0$ for $r<2m$. Using this we get in this limit
\be \label{tre27}  \frac{  \partial_u\partial_{u'} \langle in|\hat\phi(x)\hat\phi(x')|in\rangle }{(1+V(r))(1+V(r'))} |_{t=t'}=
\frac{\hbar}{4\pi (1-\sqrt{\frac{2m}{r}}\ )(1-\sqrt{\frac{2m}{r'}}\ )}\frac{1}{16m^2} \frac{1}{\cosh^2(\frac{u-u'}{8m})}|_{t=t'}\ ,  \ee
which is extremized, for $r,r'$ sufficiently away from the horizon, by $u=u'$ (i.e. along the trajectories of the particle and partner). Combining (\ref{treotto}) and (\ref{trediciannovebis}) we have
\be u= t -r - 2\sqrt{2mr} -4m\ln| \sqrt{\frac{r}{2m}}-1|\ , \label{tre28} \ee
and so the condition $u=u'$ at equal times gives 
\be \label{tre29} r+2\sqrt{2mr}+4m \ln \left( \sqrt{\frac{r}{2m}}-1\right)=r'+2\sqrt{2mr'}
+4m \ln\left(1-\sqrt{\frac{r'}{2m}} \right)\ .\ee
So in analogy with what we saw for the acoustic BH one would expect that at late time for $r,r'$ not sufficiently close to the horizon the equal time density correlator $G(t;x,x')$ should show a peak along (\ref{tre29}).
If we plot the two functions entering the left and right hand side of eq. (\ref{tre29}), as shown in Fig. (\ref{fig13}), we see the critical point explaining the apparently absurd result of our Fig. (\ref{fig10}): eq. (\ref{tre29}) has solution only for $r\lesssim 2.6m$ for which the corresponding $r'>0$. So when the Hawking particle emerges from the quantum atmosphere out of the vacuum fluctuations, at a distance $O(1/\kappa)$ (where $\kappa=\frac{1}{4m}$ is the surface gravity of the Schwarzschild black hole)  from the horizon, the corresponding partner has already been swallowed by the singularity and the correlations are lost. On the other hand, for points ($r,r'$) correlated by (\ref{tre29}) with a nonvanishing $r'$ ($<2m$) the density correlator $G(x,x')$ is dominated by the coincidence limit and, as it happens for acoustic BHs, peaks do not appear. The behaviour of $G(x,x')$  at equal times we found in Fig. (\ref{fig10}) is completely understandable. To better appreciate  the difference with the acoustic case, we have have plotted in Fig. (\ref{fig14})  the peak condition $u=u'$ for the acoustic metric eq. (\ref{duecinque}), using eq. (\ref{adieci}) at equal time. This has to be compared with Fig.(\ref{fig13}) for the gravitational BH. The appearance at equal time of the peak in the acoustic case and the non appearance in the gravitational case is self evident.


\begin{figure}[h]
\centering
\begin{subfigure}[h]{0.45\textwidth}
{\includegraphics[width=3in]{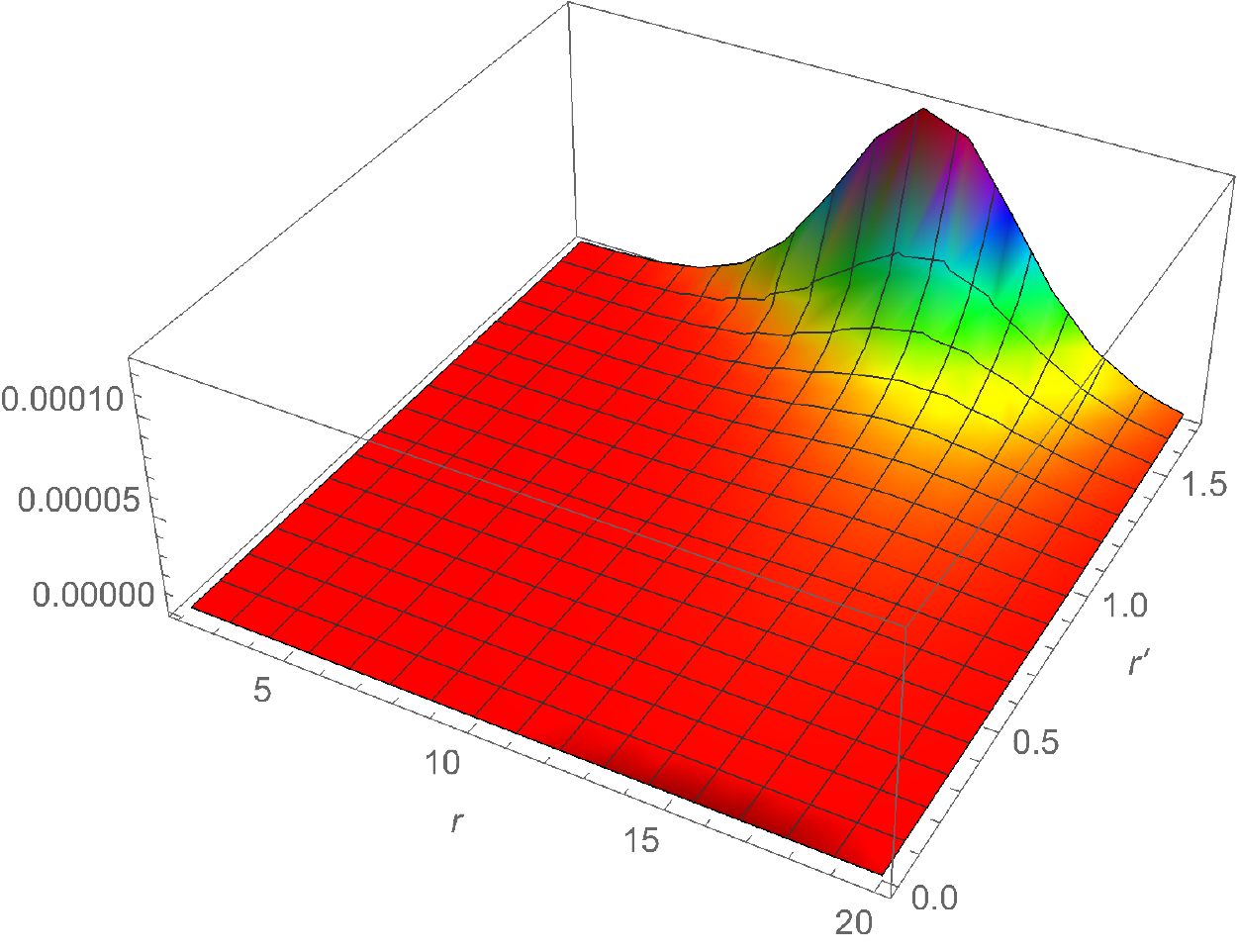}}
\caption{}
\label{15a}
\end{subfigure}
\\ \begin{subfigure}[h]{0.45\textwidth}
{\includegraphics[width=3in]{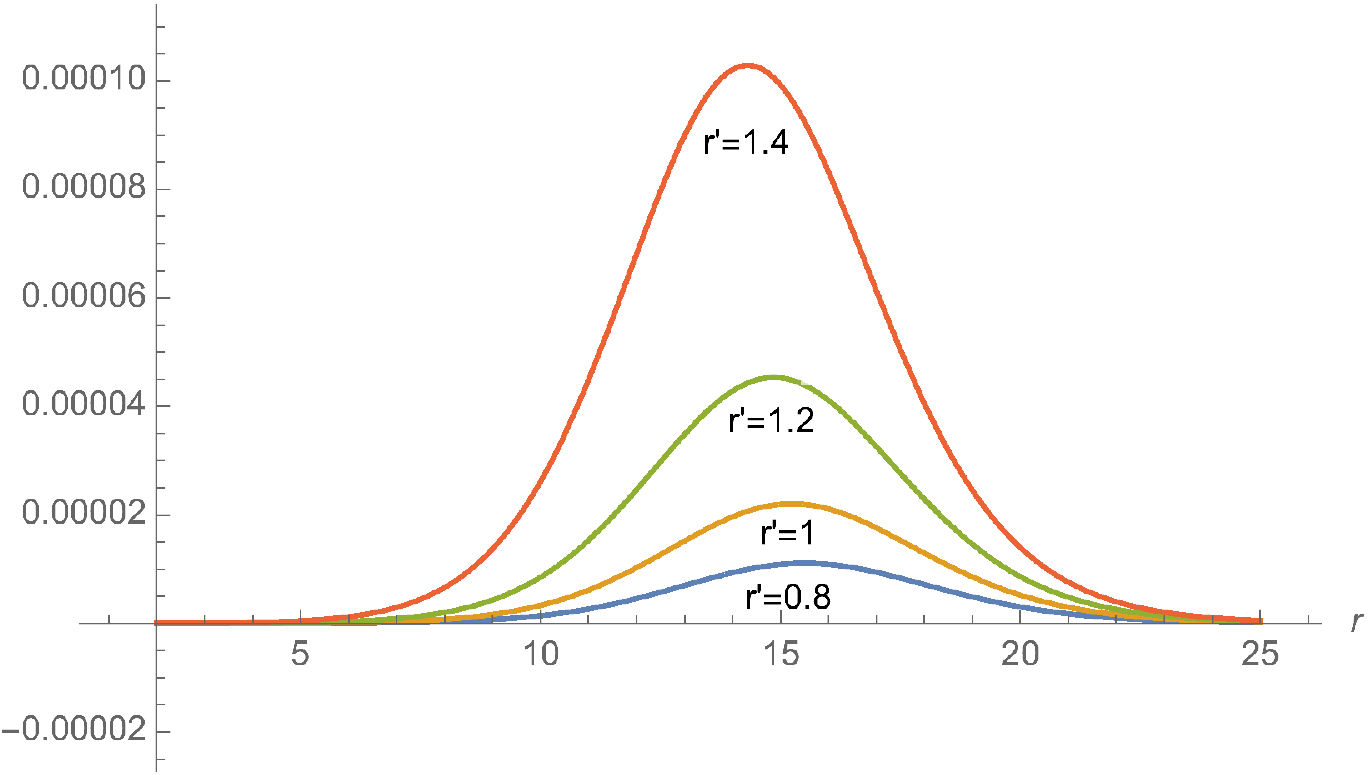}}
\caption{}
\label{15b}
\end{subfigure}
\caption{\label{fig15} a) 3D Plot of $G(x,x')$ at late time for $\Delta t =30m$. \\
b) The same correlator for fixed $r'=0.8m, m,1.2m,1.4m$. }
\end{figure}

To show the particle-partner correlation present in Hawking radiation, we have to consider the  correlator $G(x,x')$ no longer at equal time but at $t$ sufficiently bigger than $t'$ 
 so that the particle is sufficiently far away from the horizon while the partner has not yet been swallowed by the singularity.
Indeed, we see in Fig. (\ref{15a}) the emergence of a characteristic peak structure, the locus of the peak, see Fig. (\ref{15b}), being compatible with the $u=u'$ condition. 

For a condensed matter analogue of this aspect see \cite{cond-sch} where a proposal for an analogue Schwarzschild BH by using condensates of
light is made. Their Fig.4 describes the expected signals in the correlation functions at unequal times.

For completeness one has to say that if one considers just the  $\langle in| T_{uu}(x)  T_{u’u'}(x’) | in \rangle$ correlator  one finds a maximum  confined close to the horizon. This very  localised structure emerges because this correlator, unlike $G(x,x')$,  vanishes  when $x$ or $x’$ is on the horizon (see
eqs. (\ref{bundici}), (\ref{bquattordici})) for $u_{in}=0$ or $u_{in}’=0$), see also Ref. \cite{schutzholdunruh}. This is due to the pathological behaviour of the $u$ modes at the horizon.
A similar argument holds for  the $\frac{T_{uu}}{f}$ correlator.




\section{Conclusions}
In this paper we have used the methods of QFT in curved space to investigate a characteristic feature of Hawking BH radiation, namely the quantum correlation across the horizon between the
Hawking particles and their partners which should show the genuine pair creation mechanism at the origin oh this effect. The analysis concerned both acoustic BHs formed by BECs, where the
predicted correlations have indeed been experimentally observed \cite{jeff2016, jeff2019}, and the more standard BHs formed by gravitational collapse. We considered not just the late time stationary regime, but following
the time evolution of the correlations, we were able to find, in the acoustic BH case, where and when Hawking radiation appears once an horizon has formed \cite{nostro-nuovo}.
The gravitational BH led us to a puzzling result: the characteristic peak in the equal time correlation function, signaling the particle-partner correlations, we found in the acoustic case,  is here absent.
The reason for this unexpected result lies in the presence of the BH central singularity. As we have seen, Hawking particles are produced in a region displaced from the horizon, the so called "quantum atmosfere". Once the Hawking particle emerges out of this region, the corresponding partner has already been swallowed by the singularity and their mutual correlations are  lost.
Despite this negative result, we explicitly showed in Fig. (\ref{fig15})  that the Hawking quanta-partner peak does indeed show up if we consider correlators at unequal times, allowing the Hawking particle to exit the quantum atmosphere region before its partner has reached the singularity.

Finally, as already mentioned in section 2 our results do not take into account the backscattering of the modes caused by the inhomogeneities in the  metric. In the acoustic case  the effects of these inhomogeneities  are to slightly alter (up to 10 $\%$ \cite{paper2013}) the Hawking quanta-partner peak and to cause the appearance of two other signals in the correlation pattern related to the $u-v$ scattering \cite{Macher:2009nz} and significantly smaller than the main signal.  Needless to day, so far only the signal corresponding to the Hawking quanta-partner peak has been observed in a BEC acoustic BH. To discuss these two additional correlation channels for the Schwarzshild BH a full four dimensional analysis would be needed which requires heavy numerical work to get the complete two-points function
(see for example references \cite{busscasals, llos})  and which is outside the scope of the present paper.

\acknowledgments
A.F. acknowledges partial financial support by the Spanish grants FIS2017-84440-C2-1-P funded by MCIN/AEI/10.13039/501100011033 ''ERDF A way of making Europe'', Grant PID2020-116567GB-C21 funded by MCIN/AEI/10.13039/501100011033, and the project PROMETEO/2020/079 (Generalitat Valenciana).

\begin{appendix}

\section{BEC BHs: details of our mathematical construction}
\label{appendixA}

In this appendix we give the details fo the mathematical construction of our BEC BH model of section (\ref{s2}). 

In the 1+1D spacetime described by the metric
\be \label{auno} ds^2=-(c^2(t,x)-V^2)dt^2-2Vdtdx+dx^2 \ee
with the speed of sound profile given by eq. (\ref{dueuno}), we consider a massless scalar field  $\delta \hat \theta^{(2)}(t,x)$ satisfying the field equation
\be \hat \Box  \delta \hat \theta^{(2)}=0\ , \label{adue} \ee
where  $\hat\Box$ is the D'Alembert operator calculated from the above metric. 
Since every $(1+1)$ spacetime is (locally) conformally flat one can introduce a set of null coordinates $(x^+,x^-)$ so that the metric can be written as
\be \label{atre} ds^2=-C^2(x^+,x^-)dx^+dx^- \ee
and the wave equation (\ref{adue}) reduces simply to
\be \partial_{x^+}\partial_{x^-}\delta \hat \theta^{(2)}=0\ . \label{aquattro} \ee 
The field operator $\delta \hat \theta^{(2)}$ can be expanded in modes as
\be \label{acinque} \delta \hat \theta^{(2)}=\sum_\omega \ [ \hat a_\omega^+ \frac{e^{-i\omega x^+}}{\sqrt{2\pi\omega}}+  \hat a_\omega^- \frac{e^{-i\omega x^-}}{\sqrt{2\pi\omega}}+h.c.]\ . \ee
The vacuum state $|0\rangle$ is defined as $\hat a_\omega^{\pm}|0\rangle=0$ and the corresponding two-point function reads
\be \label{asei} \langle 0| \delta \hat \theta^{(2)}(x)\delta \hat \theta^{(2)}(x')|0\rangle =-\frac{\hbar}{4\pi}
\ln \Delta x^+\Delta x^- \ , \ee
where $\Delta x^{\pm}=x^{\pm}-x'^{\pm}$. In Eq. (\ref{asei}), we have omitted a diverging (irrelevant in our case) constant related to the IR divergence of our $(1+1)$ theory.  The choice of null coordinates selects the conformal vacuum. 

In our spacetime the set of null coordinates we choose for $t<0$ is
\be \label{asette} u_{in}=t-\frac{x}{c_{in}-|V|}\ , \ \ v_{in}=t+\frac{x}{c_{in}+|V|}\ . \ee
The associated vacuum state, that we denote as $|in\rangle$, represents a quantum state in which for $t<0$ there are no incoming quanta, both from left and right past null infinity. The corresponding two-point function is then
\be \label{aotto} \langle 0| \delta \hat \theta^{(2)}(t,x)\delta \hat \theta^{(2)}(t',x')|0\rangle =-\frac{\hbar}{4\pi} [ \ln(u_{in}-u_{in}')  +\ln (v_{in}-v_{in}') ]\  , \ee
where $t,t'<0$. 
The decoupling of the advanced $(v_{in})$ and retarded $(u_{in})$ sector is a consequence of the conformal invariance of the $(1+1)$ massless theory. 

\begin{figure}[h]
\includegraphics[width=5in]{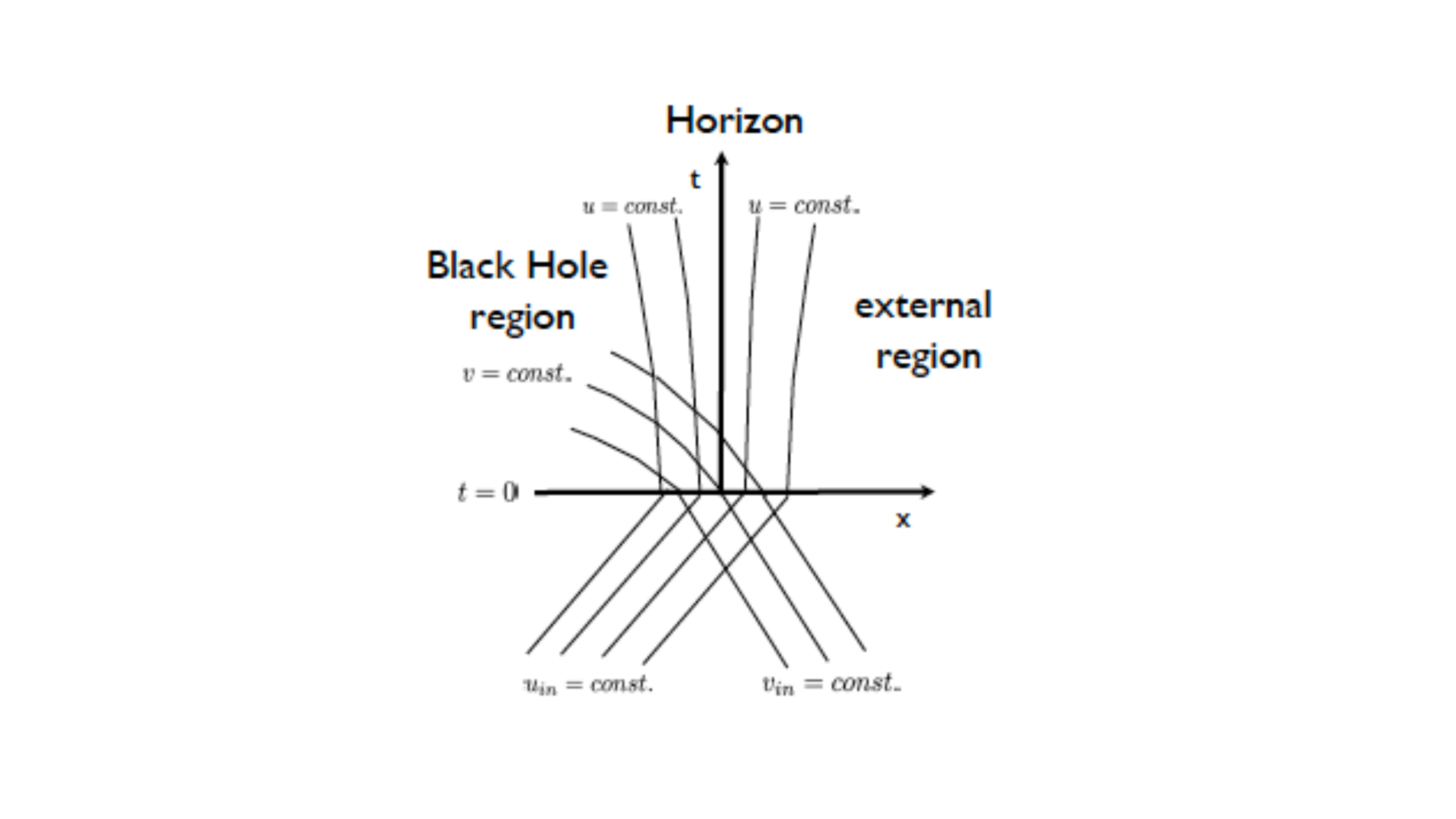} 
\caption{\label{fig16}  Null coordinates matching (in the `in' and BH regions) at the $t=0$ discontinuity.  }
\end{figure}

We have now to analytically extend the above expression in the relevant BH region, i.e. for $t,t'>0$. This is done by matching the null coordinates of eq. (\ref{asette}) with the corresponding ones in the BH region, namely
\bea u &=& t-\int \frac{dx}{c(x)-|V|}=  t-\frac{1}{\kappa}\ln\sinh \frac{3\kappa|x|}{2|V|}, \label{adieci}
\\  v&=& t+\int \frac{dx}{c(x)+|V|}= t+\frac{1}{8\kappa}\Big[\frac{9\kappa x}{2|V|}-\ln\cosh\left( \frac{3\kappa x}{2|V|}+\tanh^{-1}\frac{1}{3}\right)\Big]\ , \label{aundici}     \eea
along the discontinuity at $t=0$. See Fig. (\ref{fig16}).

Starting with the retarded coordinates, from the first of (\ref{asette}) at $t=0$ we have
\be \label{adodici} u_{in}=-\frac{x}{c_{in}-|V|}\ .\ee
Inserting this into (\ref{adieci}) evaluated at $t=0$ we get 
\be -\kappa u =\ln\sinh|Au_{in}| \label{atredici} \ , \ee
where \be \label{atredicibis} A=\frac{3\kappa}{2|V|}(|V|-c_{in})<0 \ .\ee
The relation can be inverted giving 
\be |u_{in}A|=\ln( \sqrt{1+e^{-2\kappa u}}+e^{-\kappa u})\ . \label{aquattordici} \ee 
We proceed similarly for the advanced null coordinates obtaining 
\be v_{in}=\frac{4\kappa v}{B}+\frac{1}{2B}\ln\left( 1+\sqrt{1+2\sqrt{2}e^{-8\kappa v}}\ \right)
\ ,     \label{aquindici} \ee
where \be \label{aquindicibis} B=\frac{ 3\kappa(c_{in}+|V|)}{2|V|}\ .\ee

So the two-point function for $t,t'>0$ can be written formally as 
\be \label{asedici} \langle 0| \delta \hat \theta^{(2)}(t,x)\delta \hat \theta^{(2)}(t',x')|0\rangle =-\frac{\hbar}{4\pi} \{  \ln [ u_{in}(u(t,x))-u_{in}'(u'(t',x')) ]  +\ln [ v_{in}(v(t,x))-v_{in}'(v'(t',x')) ] \} \  , \ee
where $u_{in}(u(t,x))$ is given by eqs. (\ref{aquattordici}) and (\ref{adieci}) and $v_{in}(v(t,x))$ by eqs. (\ref{aquindici}) and (\ref{aundici}). 

We have, now, all is needed to compute the density-density equal time correlator (see eq. (\ref{duedue}))

\be G_2^{(1)}(t;x,x') = \frac{n(x)n(x')}{m^2c^2(x)c^2(x')}\lim_{t\to t'} \sqrt{\frac{m^2c(x)c(x')}{n(x)n(x')}}  D  \langle  \delta \hat \theta^{(2)}(t,x)  \delta \hat \theta^{(2)}(t',x') \rangle \label{adiciassette}\ , \ee
where the differential operator $D$ is defined in eq. (\ref{duetre}), giving
\bea \label{adiciotto} 
&&G_2^{(1)}(t;x, x') = -\frac{\hbar n}{4\pi  mc(x)^{1/2}c(x')^{1/2}}\Big[ \frac{1}{(c(x)-|V|)(c(x')-|V|)} \frac{du_{in}}{du}\frac{du'_{in}}{du'}  \frac{1}{(u_{in}-u'_{in})^2} \nonumber  \\ &&
+ \frac{1}{(c(x)+|V|)(c(x')+|V|)}  \frac{dv_{in}}{dv} \frac{dv'_{in}}{dv'} \frac{1}{(v_{in}-v'_{in})^2} \Big] |_{t=t'}
\equiv G_{2,u}^{(1)}(t;x, x') +G_{2,v}^{(1)}(t;x, x'), \ \ \ \ \ \ \ \  \eea
where
\bea 
&& G_{2,u}^{(1)}(t;x, x') =  - \frac{\hbar n}{4\pi m }  \frac{\kappa^2}{|V|^3}  \frac{1}{\sqrt{(1+\frac{2}{3}\tanh \frac{3\kappa x}{2|V|})}\ \frac{2}{3} \tanh \frac{3\kappa x}{2|V|}}
  \frac{1}{\sqrt{(1+\frac{2}{3}\tanh \frac{3\kappa x'}{2|V|})}\ \frac{2}{3} \tanh \frac{3\kappa x'}{2|V|}}
  \times  \nonumber \\ && \frac{e^{-2\kappa t} \sinh\frac{3\kappa x}{2|V|} \sinh\frac{3\kappa x'}{2|V|}}{\sqrt{(1+e^{-2\kappa t} \sinh^2\frac{3\kappa x}{2|V|})(1+e^{-2\kappa t }\sinh^2\frac{3\kappa x'}{2|V|})}}
  \frac{1}
 {\left( \ln \frac{ \sqrt{1+e^{-2\kappa t}\sinh^2\frac{3\kappa x}{2|V|}}\ +\ e^{-\kappa t}\sinh \frac{3\kappa x}{2|V|}} { \sqrt{1+e^{-2\kappa t}\sinh^2\frac{3\kappa x'}{2|V|}}\ +\ e^{-\kappa t}\sinh \frac{3\kappa x'}{2|V|}} \right)^2}
\label{adiciannove}\eea
and
\bea 
&& G_{2,v}^{(1)}(t;x, x')=  - \frac{\hbar n}{4\pi m}  \frac{1}{|V|^3} \times \nonumber \\ &&  \frac{1}{\sqrt{(1+\frac{2}{3}\tanh \frac{3\kappa x}{2|V|})}(2+\frac{2}{3} \tanh \frac{3\kappa x}{2|V|})}
  \frac{1}{\sqrt{(1+\frac{2}{3}\tanh \frac{3\kappa x'}{2|V|})}(2+ \frac{2}{3} \tanh \frac{3\kappa x'}{2|V|})}\times 
\nonumber \\ &&
\frac{\sqrt{1+e^{-8\kappa t-\frac{9}{2}\frac{\kappa x}{|V|}}(2e^{\frac{3\kappa x}{2|V|}}+e^{-\frac{3\kappa x}{2|V|}})}+1+\frac{1}{2}e^{-8\kappa t-\frac{9}{2}\frac{\kappa x}{|V|}}(2e^{\frac{3\kappa x}{2|V|}}+e^{-\frac{3\kappa x}{2|V|}})}{\sqrt{1+e^{-8\kappa t-\frac{9}{2}\frac{\kappa x}{|V|}}(2e^{\frac{3\kappa x}{2|V|}}+e^{-\frac{3\kappa x}{2|V|}})} \left( 1+
\sqrt{1+e^{-8\kappa t-\frac{9}{2}\frac{\kappa x}{|V|}}(2e^{\frac{3\kappa x}{2|V|}}+e^{-\frac{3\kappa x}{2|V|}})}\right)}\times \nonumber \\ &&
\frac{\sqrt{1+e^{-8\kappa t-\frac{9}{2}\frac{\kappa x'}{|V|}}(2e^{\frac{3\kappa x'}{2|V|}}+e^{-\frac{3\kappa x'}{2|V|}})}+1+\frac{1}{2}e^{-8\kappa t-\frac{9}{2}\frac{\kappa x'}{|V|}}(2e^{\frac{3\kappa x'}{2|V|}}+e^{-\frac{3\kappa x'}{2|V|}})}{\sqrt{1+e^{-8\kappa t-\frac{9}{2}\frac{\kappa x'}{|V|}}(2e^{\frac{3\kappa x'}{2|V|}}+e^{-\frac{3\kappa x'}{2|V|}})} \left( 1+
\sqrt{1+e^{-8\kappa t-\frac{9}{2}\frac{\kappa x'}{|V|}}(2e^{\frac{3\kappa x'}{2|V|}}+e^{-\frac{3\kappa x'}{2|V|}})}\right)}\times \nonumber \\ &&
\frac{1}{\left( \frac{9}{16|V|}(x-x')-\frac{1}{8\kappa}\ln \frac{2e^{\frac{3\kappa x}{2|V|}}+e^{-\frac{3\kappa x}{2|V|}}}{2e^{\frac{3\kappa x'}{2|V|}}+e^{-\frac{3\kappa x'}{2|V|}}}
+\frac{1}{8\kappa}\ln \frac{1+ \sqrt{1+e^{-8\kappa t-\frac{9}{2}\frac{\kappa x}{|V|}}(2e^{\frac{3\kappa x}{2|V|}}+e^{-\frac{3\kappa x}{2|V|}})} } 
{1+ \sqrt{1+e^{-8\kappa t-\frac{9}{2}\frac{\kappa x'}{|V|}}(2e^{\frac{3\kappa x'}{2|V|}}+e^{-\frac{3\kappa x'}{2|V|}})} }    
 \right)^2}
\label{aventi} \eea
correspond to the $u$ channel (where Hawking radiation appears) and the $v$ channel respectively.

We can characterize a late time limit in which stationarity is achieved by 
\be \label{aventuno} e^{-\kappa t}\sinh \frac{3\kappa |x|}{2|V|}\ll1\ ,  \ee
i.e.
\be \label{aventidue} t_{late\ time}\gg \frac{1}{\kappa} \ln \sinh \frac{3\kappa |x|}{2|V|}\   \ee
and it is clear that this time depends on the point chosen. One reaches a stationary regime earlier near the horizon $(x\simeq 0)$ and later away from it. The stationary limit of eq. (\ref{adiciannove}) then reads
\bea \label{aventitre} && G_{2,u}^{(1)}(t;x, x') = \frac{\hbar}{4\pi m n_0}  \frac{\kappa^2}{|V|^3} \times \nonumber \\ && \frac{1}{\sqrt{(1+\frac{2}{3}\tanh \frac{3\kappa x}{2|V|})}\ \frac{2}{3} \tanh \frac{3\kappa x}{2|V|}}
  \frac{1}{\sqrt{(1+\frac{2}{3}\tanh \frac{3\kappa x'}{2|V|})}\ \frac{2}{3} \tanh \frac{3\kappa x'}{2|V|}}\times
  \nonumber \\ && \frac{1}{4\cosh^2 \left( \frac{1}{2}\ln \frac{ \sinh \frac{3\kappa x}{2|V|}}{\sinh \frac{3\kappa |x'|}{2|V|}}\right)} 
 \eea
while that of (\ref{aventi}) is
\bea \label{aventiquattro}  && G_{2,v}^{(1)}(t;x, x') = - \frac{\hbar n}{4\pi m|V|^3} \times \nonumber \\ &&  \frac{1}{\sqrt{(1+\frac{2}{3}\tanh \frac{3\kappa x}{2|V|})}(2+\frac{2}{3} \tanh \frac{3\kappa x}{2|V|})}
  \frac{1}{\sqrt{(1+\frac{2}{3}\tanh \frac{3\kappa x'}{2|V|})}(2+ \frac{2}{3} \tanh \frac{3\kappa x'}{2|V|})}\times \nonumber \\ && \frac{1}{\left( \frac{9}{16|V|}(x-x')-\frac{1}{8\kappa}\ln \frac{2e^{\frac{3\kappa x}{2|V|}}+e^{-\frac{3\kappa x}{2|V|}}}{2e^{\frac{3\kappa x'}{2|V|}}+e^{-\frac{3\kappa x'}{2|V|}}}\right)^2}\ .
 \eea
 
 Instead of the initial vacuum state described by $|in\rangle$ in which there are no incoming quanta, we can consider the case in which the condensate has an initial temperature $T$ so that we have an initial thermal distribution of phonons characterized by an occupation number 
 \be    N_{\omega_{u(v)}} =   \frac{1}{e^{\frac{\hbar\omega_{u(v)}}{k_BT}}-1}\  \label{aventicinque} \ , \ee
where the Doppler rescaled frequencies are 
\be    \omega_u = \frac{\omega c_{in}}{c_{in}-|V|}\ , \  \ \omega_v=\frac{\omega c_{in}}{c_{in}+|V|}\  \label{aventisette} \ee 
and $k_B$ is Boltzmann constant. The corresponding two-point function for this thermal state, that
we denote by $|T\rangle$, for the quantum field $\delta\hat \theta^{(2)}$ is \cite{vend} 
\be \langle T| \delta \hat \theta^{(2)} (t,x) \delta \hat \theta^{(2)} (t',x')|T \rangle  = -\frac{\hbar}{4\pi} \ln \frac{\sinh A_u \Delta u_{in}}{A_u} \frac{\sinh A_v \Delta v_{in}}{A_v},     \label{aventotto} \ee  
where 
\be \label{aventinove} A_{u(v)}=\frac{\pi k_B T(c_{in}\mp |V|)}{\hbar c_{in}}\ , \ee
the minus sign in (\ref{aventinove}) corresponds to $A_u$ and the plus to $A_v$; $\Delta u_{in}=u_{in}-u_{in}',\ \Delta v_{in}=v_{in}-v_{in}'$. 

The thermal equal time density-density correlator now reads 
\bea
G_{2,T}^{(1)}(t;x,x')
 &=& -\frac{\hbar n^{(1)}}{4\pi  mc(x)^{1/2}c(x')^{1/2}}
\Big[ \frac{1}{(c(x)-|V|)(c(x')-|V|)} \frac{du_{in}}{du}\frac{du'_{in}}{du'}  \frac{A_u^2}{\sinh^2 A_u (u_{in}-u'_{in})}   +  \nonumber \\ && \frac{1}{(c(x)+|V|)(c(x')+|V|)}  \frac{dv_{in}}{dv} \frac{dv'_{in}}{dv'}  \frac{A_v^2}{\sinh^2A_v (v_{in}-v'_{in})}   \Big] |_{t=t'}   \ .  \label{atrenta} \eea
Separating the $u$ and $v$ channels we have explicitly
 \bea && \label{atrentuno} G_{2,T u}^{(1)}(t;x,x')=
  - \frac{\hbar n}{4\pi m }  \frac{\kappa^2}{|V|^3}  \times  \\ && \frac{1}{\sqrt{(1+\frac{2}{3}\tanh \frac{3\kappa x}{2|V|})}\ \frac{2}{3} \tanh \frac{3\kappa x}{2|V|}}
  \frac{1}{\sqrt{(1+\frac{2}{3}\tanh \frac{3\kappa x'}{2|V|})}\ \frac{2}{3} \tanh \frac{3\kappa x'}{2|V|}} \times \nonumber  \\ && 
  \frac{e^{-2\kappa t} \sinh\frac{3\kappa x}{2|V|} \sinh\frac{3\kappa x'}{2|V|}}{\sqrt{(1+e^{-2\kappa t} \sinh^2\frac{3\kappa x}{2|V|})(1+e^{-2\kappa t }\sinh^2\frac{3\kappa x'}{2|V|})}} 
 \frac{(\frac{k_B\pi}{\hbar \beta A}\frac{(c_{in}-|V|)}{c_{in}})^2}
 {\sinh^2\left(  \frac{k_B\pi}{\hbar \beta A}\frac{(c_{in}-|V|)}{c_{in}} \ln \frac{\sqrt{1+e^{-2\kappa t}\sinh^2\frac{3\kappa x}{2|V|}}\ +\ e^{-\kappa t}\sinh \frac{3\kappa x}{2|V|}} { \sqrt{1+e^{-2\kappa t}\sinh^2\frac{3\kappa x'}{2|V|}}\ +\ e^{-\kappa t}\sinh \frac{3\kappa x'}{2|V|}} \right)} \nonumber
  \eea
and
\bea 
&& G_{2,T v}^{(1)}(t;x,x')=- \frac{\hbar n}{4\pi m }  \frac{1}{|V|^3}\times \label{atrentadue} \\ &&  \frac{1}{\sqrt{(1+\frac{2}{3}\tanh \frac{3\kappa x}{2|V|})}(2+\frac{2}{3} \tanh \frac{3\kappa x}{2|V|})}
  \frac{1}{\sqrt{(1+\frac{2}{3}\tanh \frac{3\kappa x'}{2|V|})}(2+ \frac{2}{3} \tanh \frac{3\kappa x'}{2|V|})}\times 
\nonumber \\ &&
\frac{\sqrt{1+e^{-8\kappa t-\frac{9}{2}\frac{\kappa x}{|V|}}(2e^{\frac{3\kappa x}{2|V|}}+e^{-\frac{3\kappa x}{2|V|}})}+1+\frac{1}{2}e^{-8\kappa t-\frac{9}{2}\frac{\kappa x}{|V|}}(2e^{\frac{3\kappa x}{2|V|}}+e^{-\frac{3\kappa x}{2|V|}})}{\sqrt{1+e^{-8\kappa t-\frac{9}{2}\frac{\kappa x}{|V|}}(2e^{\frac{3\kappa x}{2|V|}}+e^{-\frac{3\kappa x}{2|V|}})} \left( 1+
\sqrt{1+e^{-8\kappa t-\frac{9}{2}\frac{\kappa x}{|V|}}(2e^{\frac{3\kappa x}{2|V|}}+e^{-\frac{3\kappa x}{2|V|}})}\right)}\times \nonumber \\ &&
\frac{\sqrt{1+e^{-8\kappa t-\frac{9}{2}\frac{\kappa x'}{|V|}}(2e^{\frac{3\kappa x'}{2|V|}}+e^{-\frac{3\kappa x'}{2|V|}})}+1+\frac{1}{2}e^{-8\kappa t-\frac{9}{2}\frac{\kappa x'}{|V|}}(2e^{\frac{3\kappa x'}{2|V|}}+e^{-\frac{3\kappa x'}{2|V|}})}{\sqrt{1+e^{-8\kappa t-\frac{9}{2}\frac{\kappa x'}{|V|}}(2e^{\frac{3\kappa x'}{2|V|}}+e^{-\frac{3\kappa x'}{2|V|}})} \left( 1+
\sqrt{1+e^{-8\kappa t-\frac{9}{2}\frac{\kappa x'}{|V|}}(2e^{\frac{3\kappa x'}{2|V|}}+e^{-\frac{3\kappa x'}{2|V|}})}\right)}\times \nonumber \\ &&
\frac{(\frac{4\kappa}{B}\frac{k_B \pi}{\hbar \beta }\frac{(c_{in}+|V|)}{c_{in}})^2}{\sinh^2 \left(  \frac{4\kappa}{B}\frac{k_B \pi}{\hbar \beta }\frac{(c_{in}+|V|)}{c_{in}} \Big( \frac{9}{16|V|}(x-x')-\frac{1}{8\kappa}\ln \frac{2e^{\frac{3\kappa x}{2|V|}}+e^{-\frac{3\kappa x}{2|V|}}}{2e^{\frac{3\kappa x'}{2|V|}}+e^{-\frac{3\kappa x'}{2|V|}}}
+\frac{1}{8\kappa}\ln \frac{1+ \sqrt{1+e^{-8\kappa t-\frac{9}{2}\frac{\kappa x}{|V|}}(2e^{\frac{3\kappa x}{2|V|}}+e^{-\frac{3\kappa x}{2|V|}})} } 
{1+ \sqrt{1+e^{-8\kappa t-\frac{9}{2}\frac{\kappa x'}{|V|}}(2e^{\frac{3\kappa x'}{2|V|}}+e^{-\frac{3\kappa x'}{2|V|}})} }    
 \Big)\right)}. \nonumber
\eea
Note that by introducing the Hawking temperature $T_H=\frac{\hbar \kappa}{2\pi k_B}$ we can rewrite the prefactor of the $\ln $ in eq. (\ref{atrentuno}), see also (\ref{atredicibis}), as $\frac{k_B\pi}{\hbar \beta A}\frac{(c_{in}-|V|)}{c_{in}}=-\frac{T}{T_H}\frac{|V|}{3c_{in}}$, and, in (\ref{atrentadue}), see also (\ref{aquindicibis}), 
$\frac{4\kappa}{B}\frac{k_B \pi}{\hbar \beta}\frac{(c_{in}+|V|)}{c_{in}}=\frac{T}{T_H}\frac{4|V|\kappa}{3c_{in}}$.   We see that stationarity of $G_{2,T u}^{(1)}$ is reached when $T>\frac{3c_{in}}{|V|}T_H$ not just by the previous condition  $e^{-\kappa t}\sinh \frac{3\kappa |x|}{2|V|}\ll1$, which holds for $T<\frac{3c_{in}}{|V|}T_H$, but by the most stringent conditon 
\be \frac{T}{ \frac{3c_{in}}{|V|}T_H}e^{-\kappa t}\sinh \frac{3\kappa |x|}{2|V|}\ll1\ .\label{atrentatre} \ee
The denominator of the last term in (\ref{atrentuno}) can be approximated as follows
\bea \label{atrentaquattro} && \sinh^2\left(  \frac{k_B\pi}{\hbar \beta A}\frac{(c_{in}-|V|)}{c_{in}} \ln \frac{ \sqrt{1+e^{-2\kappa t}\sinh^2\frac{3\kappa x}{2|V|}}\ +\ e^{-\kappa t}\sinh \frac{3\kappa x}{2|V|}} { \sqrt{1+e^{-2\kappa t}\sinh^2\frac{3\kappa x'}{2|V|}}\ +\ e^{-\kappa t}\sinh \frac{3\kappa x'}{2|V|}} \right)
\\ && \simeq \sinh^2\left(  \frac{k_B\pi}{\hbar \beta A}\frac{(c_{in}-|V|)}{c_{in}} e^{-\kappa t}(\sinh \frac{3\kappa x}{2|V|}-\sinh \frac{3\kappa x'}{2|V|})\right)\nonumber \\ && \simeq  \left(\frac{k_B\pi}{\hbar \beta A}\frac{(c_{in}-|V|)}{c_{in}} \right)^2e^{-2\kappa t}\left( \sinh \frac{3\kappa x}{2|V|}-\sinh \frac{3\kappa x'}{2|V|} \right)^2\ .  \nonumber  \eea
Inserting this in eq. (\ref{atrentuno}) and approximating  $\sqrt{1+e^{-2\kappa t}\sinh^2\frac{3\kappa x}{2|V|}}\simeq 1$ we have 
 \bea G_{2,T u}^{(1)}(t;x,x') &=&
  - \frac{\hbar n}{4\pi m }  \frac{\kappa^2}{|V|^3}  \frac{1}{\sqrt{(1+\frac{2}{3}\tanh \frac{3\kappa x}{2|V|})}\ \frac{2}{3} \tanh \frac{3\kappa x}{2|V|}}
  \frac{1}{\sqrt{(1+\frac{2}{3}\tanh \frac{3\kappa x'}{2|V|})}\ \frac{2}{3} \tanh \frac{3\kappa x'}{2|V|}} \times\ \nonumber
 \\ && \frac{ \sinh\frac{3\kappa x}{2|V|} \sinh\frac{3\kappa x'}{2|V|}}
 {(\sinh \frac{3\kappa x}{2|V|}-\sinh \frac{3\kappa x'}{2|V|})^2}  \label{atrentacinque} \eea
which (for points $x,x'$ situated on both sides of the horizon) is exactly the same as the one at $T=0$, see eq. (\ref{aventitre}). One can for completeness evaluate the late time limit of $G_{2,T v}^{(1)}$
giving 
\bea 
&& G_{2,T v}^{(1)}(t;x,x')=- \frac{\hbar n}{4\pi m }  \frac{1}{|V|^3}\times   \nonumber \\ && \frac{1}{\sqrt{(1+\frac{2}{3}\tanh \frac{3\kappa x}{2|V|})}(2+\frac{2}{3} \tanh \frac{3\kappa x}{2|V|})}
  \frac{1}{\sqrt{(1+\frac{2}{3}\tanh \frac{3\kappa x'}{2|V|})}(2+ \frac{2}{3} \tanh \frac{3\kappa x'}{2|V|})} 
  \times
\nonumber \\ &&  \frac{(\frac{4\kappa}{B}\frac{\pi}{\beta }\frac{(c_{in}+|V|)}{c_{in}})^2}{\sinh^2 \left(  \frac{4\kappa}{B}\frac{\pi}{\beta }\frac{(c_{in}+|v|)}{c_{in}} \Big( \frac{9}{16|V|}(x-x')-\frac{1}{8\kappa}\ln \frac{2e^{\frac{3\kappa x}{2|v|}}+e^{-\frac{3\kappa x}{2|V|}}}{2e^{\frac{3\kappa x'}{2|v|}}+e^{-\frac{3\kappa x'}{2|V|}}}   
 \Big)\right)}\ . \label{atrentasei} \eea
 
 \section{Gravitational BHs: details of our mathematical construction}

\label{appendixB}

The Schwarzschild metric in Painlev\'e coordinates is (we consider the two dimensional section) 
\be \label{buno} ds^2=-fdt^2-2Vdtdr+dr^2\ .\ee
The components of the four velocity of a geodesic observer free falling from infinity with initial zero velocity are
\be \label{bdue} u^a=(1,V)\ .\ee
So
\be \label{btre} \hat T_{ab}u^au^b=\hat T_{tt} +2V\hat T_{rt}+V^2\hat T_{rr}\ . \ee
The vanishing of the trace of $\hat T_{ab}$ gives 
\be \label{bquattro} g^{ab}\hat T_{ab}=-\hat T_{tt}-2V\hat T_{tr}+f\hat T_{rr}=0\ee
which inserted in eq. (\ref{btre}) gives 
\be \label{bcinque} \hat T_{ab}u^au^b=\hat T_{rr} \ . \ee
Given the form of the two-point function (see eq. (\ref{trequindici})) it is useful for the calculations to express $\hat T_{rr}$ in terms of $\hat T_{uu}$ and $\hat T_{vv}$ (note that $\hat T_{uv}=0$) where $u,v$ are Eddington-Finkelstein null coordinates. Performing the coordinate transformation we have 
\be \label{bsei} \hat T_{rr}(t,r)=\Big[ \frac{\sqrt{1-f} +1}{f}\Big]^2 \hat T_{uu}(u,v)+ 
\Big[ \frac{\sqrt{1-f} -1}{f}\Big]^2 \hat T_{vv}(u,v)\ , \ee
which taking into account that $f=1-V^2$, $V<0$, yields
\be \label{bsette} \hat T_{rr}= \frac{\hat T_{uu}}{(1+V)^2}+ \frac{\hat T_{vv}}{(1-V)^2}\ , \ee
where $V=-\sqrt{\frac{2m}{r}}$. 

From (\ref{trediciassette}) we have 
\be \label{botto} \hat T_{uu}=\partial_u \hat \phi \partial_u \hat \phi \ee
and similarly
\be \label{bnove} \hat T_{vv}=\partial_v \hat \phi \partial_v \hat \phi \ . \ee
The fundamental object for our calculation is
\be \label{bdieci}  \partial_u \partial_{u'} \langle in| \hat\phi(x)\hat \phi(x')|in\rangle=-\frac{\hbar}{4\pi}
\frac{du_{in}}{du}\frac{du_{in}'}{du'}\frac{1}{(u_{in}-u_{in}')^2} \ee
where use of the two-point function (\ref{trequindici}) was made. From this the correlator of 
$\hat T_{uu}$ reads formally
\be \label{bundici} \langle in|\hat T_{uu}(x)\hat T_{u'u'}(x')|in\rangle =
(\partial_u\partial_{u'} \langle in|\hat\phi(x)\hat\phi(x')|in\rangle)^2 \ . \ee
For the $v$ sector
\be \label{bdodici}  \partial_v \partial_{v'} \langle in| \hat\phi(x)\hat \phi(x')|in\rangle=-\frac{\hbar}{4\pi}
\frac{1}{(v-v')^2}\ ,  \ee
so
\be \label{btredici} \langle in|\hat T_{vv}(x)\hat T_{v'v'}(x')|in\rangle =
\frac{\hbar}{(4\pi)^2}\frac{1}{(v-v')^4}  \ . \ee
Taking into account the relation between $u$ and $u'$ (see eq. (\ref{tredodici})) we can rewrite (\ref{bdieci}) as
\bea \label{bquattordici} 
 \partial_u \partial_{u'} \langle in| \hat\phi(x)\hat \phi(x')|in\rangle &=& -\frac{\hbar}{4\pi}
 \frac{u_{in}u_{in}'}{(u_{in}-u_{in}')^2}\frac{1}{(u_{in}-4m)(u_{in}'-4m)}\\
 &=&-\frac{\hbar}{4\pi}\frac{1}{4\cosh^2\ln\sqrt{\frac{-u_{in}}{u_{in}'}}}\frac{1}{(u_{in}-4m)(u_{in}'-4m)}
 \ , \nonumber  \eea
 where we have chosen the point $x$ outside the horizon ($u_{in}<0$) and $x'$ inside the horizon ($u_{in}'>0$) [more precisely, see eq. (\ref{tretredici})], 
 \be \label{bquindici} u_{in}=-4mW(e^{-\frac{u}{4m}}),\ \ u_{in}'=-4mW(-e^{-\frac{u'}{4m}})\ . \ee
 We see that the $v$ sector gives just vacuum correlations. The correlations in Hawking radiation come only from the $u$ sector, this is a consequence of the conformal invariance of our model which does not take into account backscattering of the modes of the quantum field. 
The mixed correlator can be similarly computed 
\be \label{bsedici} \langle in|\hat T_{uu}(x)\hat T_{v'v'}(x')|in\rangle = 2(\partial_u\partial_{v'}\langle in|\hat \phi(x)\hat\phi(x')|in\rangle)^2\ , \ee 
\be \label{bdiciassette} \langle in|\hat T_{vv}(x)\hat T_{u'u'}(x')|in\rangle = 2(\partial_v\partial_{u'}\langle in|\hat \phi(x)\hat\phi(x')|in\rangle)^2\ , \ee 
with 
\be \label{bdiciotto} \partial_u\partial_{v'}\langle in|\hat\phi(x)\hat\phi(x')|in\rangle=-\frac{\hbar}{4\pi}\frac{u_{in}}{(u_{in}-4m)}\frac{1}{(u_{in}-v')^2}\  \ee
and     
\be \label{bdiciannove} \partial_v\partial_{u'}\langle in|\hat\phi(x)\hat\phi(x')|in\rangle=-\frac{\hbar}{4\pi}\frac{u_{in}'}{(u_{in}'-4m)}\frac{1}{(u'_{in}-v)^2}\ . \ee
     
\end{appendix}

\end{document}